%% file: HK_Zerspanungsversuche.tex
\documentclass[preprint,12pt]{elsarticle}

\usepackage[a4paper]{geometry}													
\usepackage{amssymb}
\usepackage{caption}
\usepackage{amsmath}
\usepackage{color}																	
\usepackage{colortbl}																
\usepackage{multirow}
\usepackage{longtable}			
\usepackage{hyperref}
\hypersetup{					
    colorlinks=true,
    linkcolor=blue,
    filecolor=magenta,      
    urlcolor=cyan,
    pdftitle={Dry Cutting Experiments Database},
    }
\usepackage{svg}

\usepackage{siunitx}

\usepackage{placeins}			

\newcommand{\Titan}{\TitanPunkt{ }}	
\newcommand{\TitanPunkt}{Ti6Al4V}	

\makeatletter
\def\ps@pprintTitle{%
    \let\@oddhead\@empty
	\let\@evenhead\@empty
	\def\@oddfoot{\reset@font\hfil\thepage\hfil}
	\let\@evenfoot\@oddfoot
}
\makeatother

\begin{document}

\begin{frontmatter}



\title{Dry Cutting Experiments Database \Titan and Ck45}


\author{Hagen Klippel*}
\ead{klippel@iwf.mavt.ethz.ch}

\author{Hagen Klippel, Stefan S\"ussmaier, Michal Kuffa, Konrad Wegener}

\address{Institute of Machine Tools and Manufacturing (IWF), Department of Mechanical and Process Engineering, ETH Z\"urich, Leonhardstrasse 21, 8092 Z\"urich, Switzerland}

\begin{abstract}

The numerical simulation of metal cutting processes requires material data for constitutive equations, which cannot be obtained with standard material testing procedures. Instead, inverse identifications of material parameters within numerical simulation models of the cutting experiment itself are necessary. The intention of the present report is the provision of results of a large scale experimental study of dry orthogonal cutting experiments of \Titan (3.7165 Grade 5) and Ck45 (AISI 1045) along with their documentation and interpretation. The process forces are evaluated and each cutting insert geometry has been measured prior to the experiments to determine the cutting edge radii for each experiment. The resulting chip forms are analysed and the averaged chip thicknesses are determined. A material characterization is performed, which includes microstructural investigations on the raw materials and is reported together with tensile test results. The assembled data set can be used for parameter identification when the experimental conditions are reproduced in numerical simulations. The cutting test results are finally used to derive coefficients for Kienzle's force model. The data is stored in the \href{https://my.pcloud.com/}{pCloud} and contains process force measurement data, cutting edge radii scans, pictures of chip geometries and etched chips.

\end{abstract}


\end{frontmatter}

\section{Introduction}
\label{Kap:Introduction}

The metal cutting process is characterized by harsh conditions, which comprise large plastic strains up to $700\%$, strain rates up to $10^6s^{-1}$ and temperatures ranging from $500^\circ C$ to $1400^\circ C$, according to \cite{Arrazola2013} . In the numerical simulation of such processes material parameters for constitutive models are required, but these conditions cannot be reproduced in neither simple (direct) experiments nor in experiments combining these effects. Instead, material parameters need to be deduced from an inverse identification, where the cutting test itself serves as a material test, as for example shown in 
\cite{Hardt2021,Karandikar2022,Thimm2018}.


Constitutive model constants taken from literature, e.g. for \Titan \cite{Ducobu2017}, show large variations and so do the simulation results. Raw materials specified by the respective norms, when sourced from different suppliers and charges, shows slight variations in the chemical composition within the tolerance bands and may have undergone different manufacturing steps (drawing, rolling, forging) and heat treatments, which finally leads to a wide scatter of thermo-mechanical behaviour \cite{Hokka2018,Luttervelt1998}. These initial conditions of the materials are often not clear, which makes it difficult to select a suitable constitutive model parameter set from literature for the numerical simulation of documented experiments, for example in \cite{Ducobu2015b,Wyen2011}.



Ck45 is a perlitic-ferritic steel which is easy to machine and widely used in mechanical engineering applications, e.g. for crankshafts, bolts, gears and bearings \cite{Otai}. The material exhibits dynamic strain aging at elevated temperatures leading to increased yield stresses depending on plastic strain rate and temperature and thus can affect the chip formation and process forces as highlighted in \cite{Childs2019,Devotta2015,Devotta2020}.

Titanium alloys are of great industrial interest \cite{Peters2002} and find wide use use in biotechnical applications due to their biocompatibility \cite{Froes2018} and are also popular in aeronautics \cite{Peters2003} due to their low density and high strength. The most widely used titanium alloy is \Titan which however is considered difficult to machine because the material removal rate is low if the tool wear is to be low, since the considerable high-temperature strength inhibits temperature-induced softening during the separation process, leading to high thermal and mechanical stress in the tool \cite{Jaffery2009,Pramanik2015,Younas2021}.

The aim of the present publication is the documentation of a large scale testing program of (AISI 1045) and \Titan (3.7165, Grade 5) using dry quasi-orthogonal cutting experiments. The cutting tests are performed in a wide range of feed rates ($f=0.01..0.4 mm/rev$) and cutting speeds ($v_c=10...500 m/min$). Prior to the cutting experiments the raw materials are examined: hardness tests, as well as microstructural analyses by means of investigating etched samples and electron backscatter diffraction (EBSD) analyses are conducted to detect any irregularities, e.g. preferential grain orientations or material anisotropies, that may be caused by the manufacturing process. Tensile tests are carried out on test specimens of these materials and material parameter for quasi-stationary conditions together with the rate dependency at low strain rates, are documented.

In the orthogonal cutting tests neither lubrication nor cooling have been used. Each cutting test was performed with an unused cutting edge to minimize possible wear effects on the process. All cutting edges were measured using 3D metrology before the cutting tests to determine the cutting edge radii in the unworn state. This is necessary because the cutting edge radii can have a significant influence the process forces \cite{Albrecht1960,Wyen2010,Wyen2011}. The measured process forces, chip shapes, average chip thicknesses, chip microstructures, apparent friction coefficients according to Merchant \cite{Merchant1945a} and where applicable built-up edge formations and tempering colors are documented. The measured process forces are used to derive Kienzle coefficients for both materials over a large range of cutting speeds.

\section{Material Specification}
\label{Kap:RohmaterialVersuche}

Raw materials of Ck45 and \Titan are used in cylindrical form with a diameter of $\approx \SI{80}{\milli\meter}$ and a height of $\SI{90}{\milli\meter}$. All cylinders are from the same batch of the respective material.

\subsection{Delivery Condition of the Materials}

\Titan is an alloy containing 6\% (weight) Aluminium and 4\% (weight) of Vanadium and consists of two phases: $\alpha$-phase and $\beta$-phase. The $\alpha$-phase is stabilized by aluminium and has a hcp-structure, while the $\beta$-phase is stabilized by vanadium, consisting of a bcc-lattice \cite{Babu2013}. This batch of material was produced using the triple Vacuum Arc Remelting (VAR) method. After production a heat treatment at $750^\circ C$ for $\SI{90}{\min}$ was performed followed by air cooling. The chemical composition as specified from the supplier is given in table \ref{Tab:Ti6Al4V_Ck45_ChemischeZusammensetzungLieferschein} and the tensile properties (minimum values) in table \ref{Tab:Ti6Al4VCk45_ZugtestLieferschein}.

The material batch of Ck45 (C45E) was produced in an electric shaft furnace and afterwards rolled into cylindric form. The chemical composition as specified by the supplier is shown in table \ref{Tab:Ti6Al4V_Ck45_ChemischeZusammensetzungLieferschein}, the tensile properties of the materials in table \ref{Tab:Ti6Al4VCk45_ZugtestLieferschein}. A Jominy-test was performed by the supplier and the results are shown in table \ref{Tab:Ck45_StirnabschreckversuchLieferschein}. The material exhibits a high hardenability at the outer surface where the hardness is up to 59 HRC which indicates a martensitic microstructure and decreases to 19 HRC with increasing distance from the surface, indicating a ferritic-perlitic microstructure.

\begin{table}[h]
	\scriptsize
	\setlength\tabcolsep{1 pt} 
	\begin{center}
		\begin{tabular}{ c | c | c | c | c | c | c | c | c | c | c | c | c | c | c | c | c | c | c | c | c | c | c}
			Material & Fe & C & N & H & O & Y & Al & V & Ti & Mn & Si & P & S & Cr & Ni & Mo & Cu & Sn & Nb & B & \multicolumn{2}{c}{Residuals}\\
			& & & & & & & & & & & & & & & & & & & & & each & total\\
			\hline
			\Titan & 0.111 & 0.025 & 0.020 & 0.003 & 0.15 & $<\!0.005$ & 6.12 & 4.11 & \tiny{Balance} & - & - & - & - & - & - & - & - & - & - & - & $<\!0.1$ & $<\!0.4$ \\ %
			Ck45 & \tiny{Balance} & 0.445 & - & - & - & - & 0.01 & 0.002 & 0.01 & 0.76 & 0.24 & 0.018 & 0.02 & 0.18 & 0.05 & 0.01 & 0.15 & 0.006 & 0.001 & 0.000 & - & - \\
		\end{tabular}
		\caption{\Titan and Ck45: Supplier information on the chemical composition of this material batch.}
		\label{Tab:Ti6Al4V_Ck45_ChemischeZusammensetzungLieferschein}
	\end{center}
\end{table}

\begin{table}[h]
	\footnotesize
	\begin{center}
		\begin{tabular}{ c | c | c | c | c | c}
			Material & Tensile & Yield strength & Elongation & Reduction & Hardness\\
			& strength [MPa] & (0,2\% offset) [MPa] & at break [\%] & of area [\%] & test [HRC]\\
			\hline
			\Titan & 952 & 869 & 16.5 ($4 \cdot D$, \cite{ASTM_E8_E8M_13a_2013}) & 39 & 30.0\\
			Ck45 & 671 & 420 & 22.2 ($A_5$, \cite{DIN6892_2009}) & \\
		\end{tabular}
		\caption{\Titan and Ck45: Supplier information on tensile test results of the respective material batch.}
		\label{Tab:Ti6Al4VCk45_ZugtestLieferschein}
	\end{center}
\end{table}

%
%
	
\begin{table}[h]
	\footnotesize
	\begin{center}
		\begin{tabular}{ c | c | c | c | c | c | c | c | c | c | c | c | c | c | c | c | c }
			[mm] & 1 & 2 & 3 & 4 & 5 & 6 & 7 & 8 & 9 & 10 & 11 & 13 & 15 & 20 & 25 & 30 \\
			\hline
			HRC & 59 & 57 & 54 & 46 & 37 & 34 & 30 & 29 & 28 & 27 & 26 & 25 & 24 & 23 & 21 & 19 \\
		\end{tabular}
		\caption{Ck45: supplier information on Jominy test results of this material batch.}
		\label{Tab:Ck45_StirnabschreckversuchLieferschein}
	\end{center}
\end{table}
		


\subsection{Sample Preparation}

\begin{figure}[h]
	\begin{center}
		\includegraphics[width=0.5\textwidth]{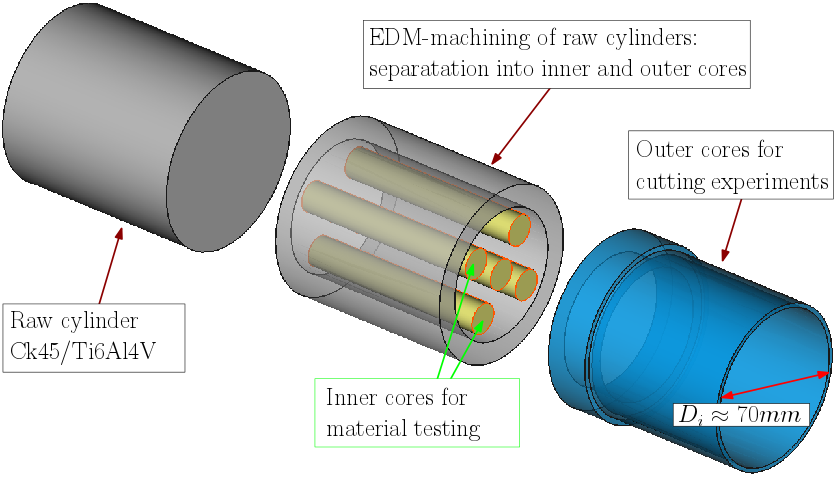}
		\includegraphics[width=0.4\textwidth]{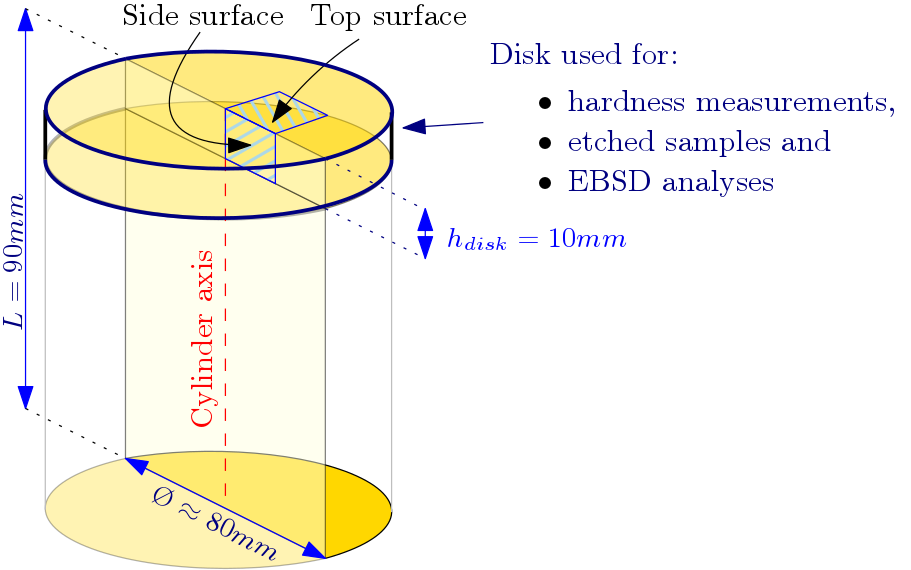}
	\end{center}
	\caption{Left: raw material usage for material tests and orthogonal cutting tests. Right: sample orientations for hardness measurements, etched samples and EBSD analysis with surface denominations.}
	\label{Bild:Rohversuchswerkstoff}
\end{figure}

Deep bores are inserted to the raw cylinders using EDM drilling. A smaller cylinder of $\SI{60}{\milli\meter}$ diamater is cut out via wire EDM and used to manufacture the test specimens for tensile testing, see chapter \ref{Kap:ZugversucheIVP}. The outer cylinders are used for orthogonal cutting experiments. A disk with a height of $h_{disk} = \SI{10}{\milli\meter}$ is sliced off of one cylinder of each material using wire EDM and used to determine the radial hardness profile. The disks are ground and polished prior to the hardness measurements. After the hardness measurements, the same disks are used to prepare etched samples from the top and side surface for microstructural investigations and EBSD analyses. An overview of the material usage is shown in Figure \ref{Bild:Rohversuchswerkstoff}.

\FloatBarrier

\subsection{Vickers Hardeness Measurement}

\begin{figure}[h]
	\begin{center}
		\includegraphics[height=0.25\textwidth]{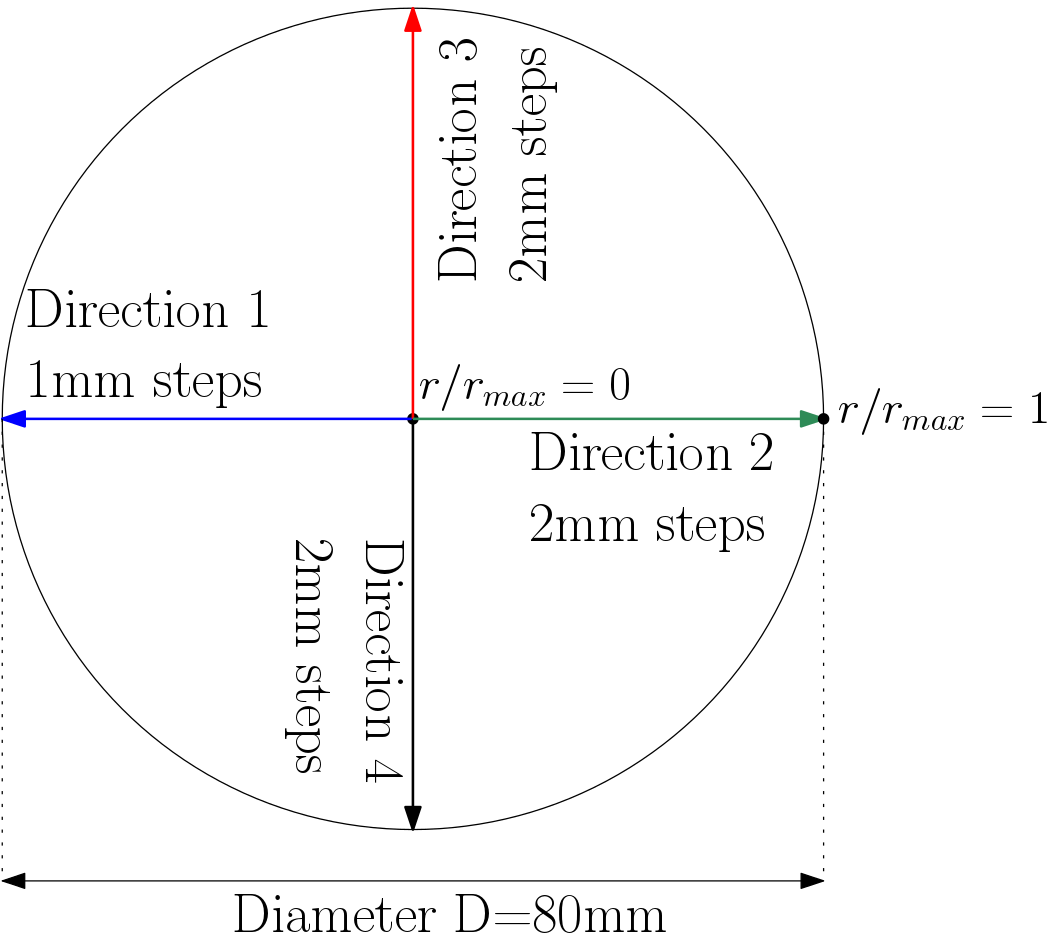}
		\includegraphics[height=0.25\textwidth]{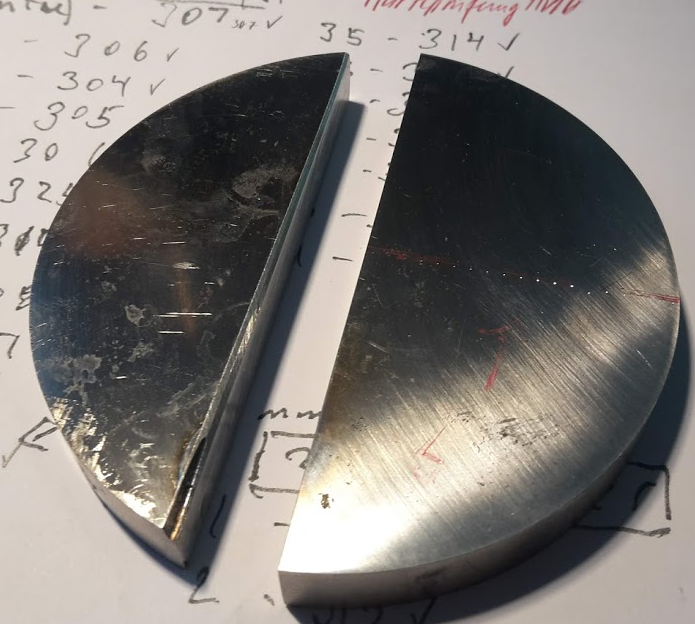}
		\includegraphics[height=0.25\textwidth]{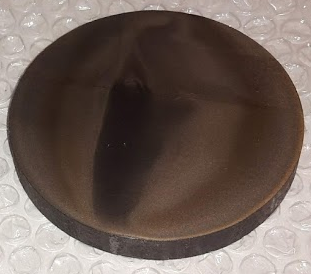}
	\end{center}
	\caption{Hardness measurement directions (left), \Titan disk (middle) and Ck45 (right) disks from cylinders used for the hardness measurements. The Ck45 disk is shown here before grinding, polishing and hardness measurements while the \Titan is already cut in half for microstructural investigations with the right half showing imprints from the hardness measurements (2mm stepping) along the slightly visible red line.}
	\label{Bild:Haertemessung_HV}
\end{figure}

\begin{figure}[h]
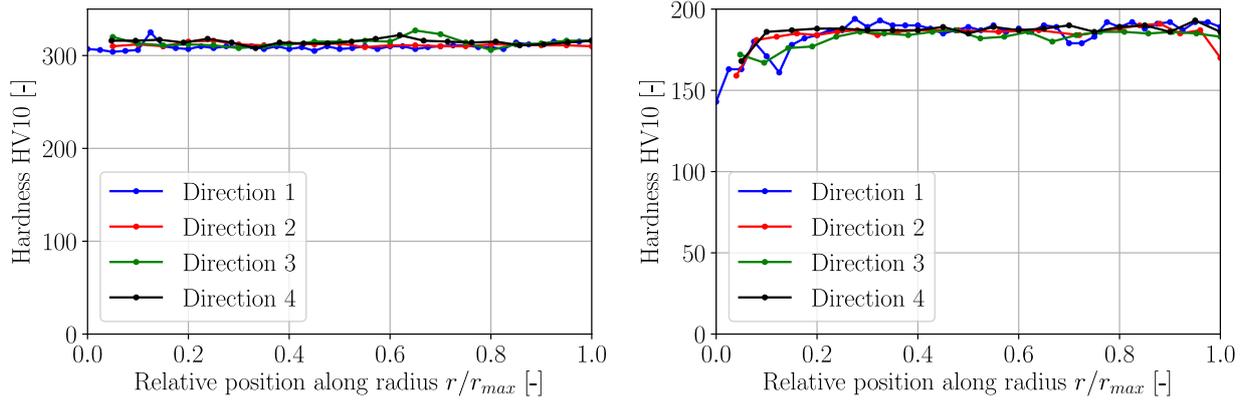

	\begin{center}
		\includesvg[width=0.49\textwidth]{Bilder/Haerteverlauf_Ti6Al4V}
		\includesvg[width=0.49\textwidth]{Bilder/Haerteverlauf_Ck45}
	\end{center}
	\caption{Hardness measurement \TitanPunkt (left) and Ck45 (right).}
	\label{Bild:Haerteverlauf_Ck45_Ti6Al4V}
\end{figure}

Vickers hardness measurements HV10 are conducted for Ck45 and \TitanPunkt. The measurements are performed on the top faces along four directions of the samples where in one direction a $\SI{1}{\milli\meter}$ stepping and in the others a $\SI{2}{\milli\meter}$ stepping was used. The hardness measurement directions and both disks are shown in figure \ref{Bild:Haertemessung_HV}. The results of the hardness measurement are shown in Figure \ref{Bild:Haerteverlauf_Ck45_Ti6Al4V} where the Ck45 shows a hardness reduction at the disk center within a radius of around 6mm. Towards the outer radius the hardness is constant except for the very last measurement point in direction 2 which shows a slight drop in hardness. The hardness distribution of the \Titan sample is almost constant in all directions and radial positions with the exception of some spots.


\FloatBarrier

\subsection{Microstructure}

The microstructure of the two materials is investigated by optical analysis of the etched surface and EBSD-analysis. This investigation should indicate possible irregularities of the grain structure or anisotropy in the material which could lead to different requirements for the constitutive model to be used in numerical simulations of the cutting experiments.

\subsubsection{Etching}

\begin{figure}[b!]
	\center{
		\includegraphics[width=0.49\textwidth]{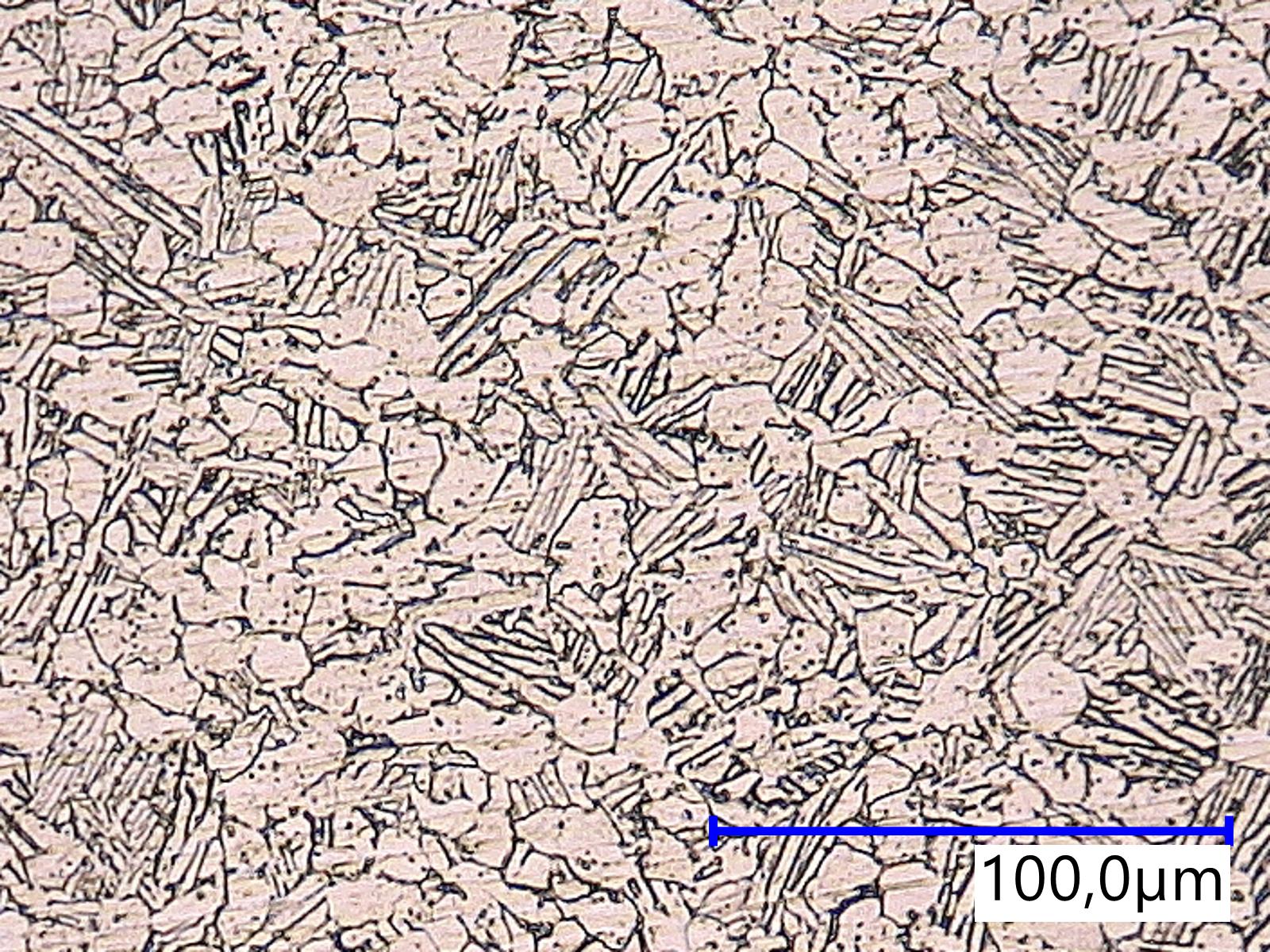}
		\includegraphics[width=0.49\textwidth]{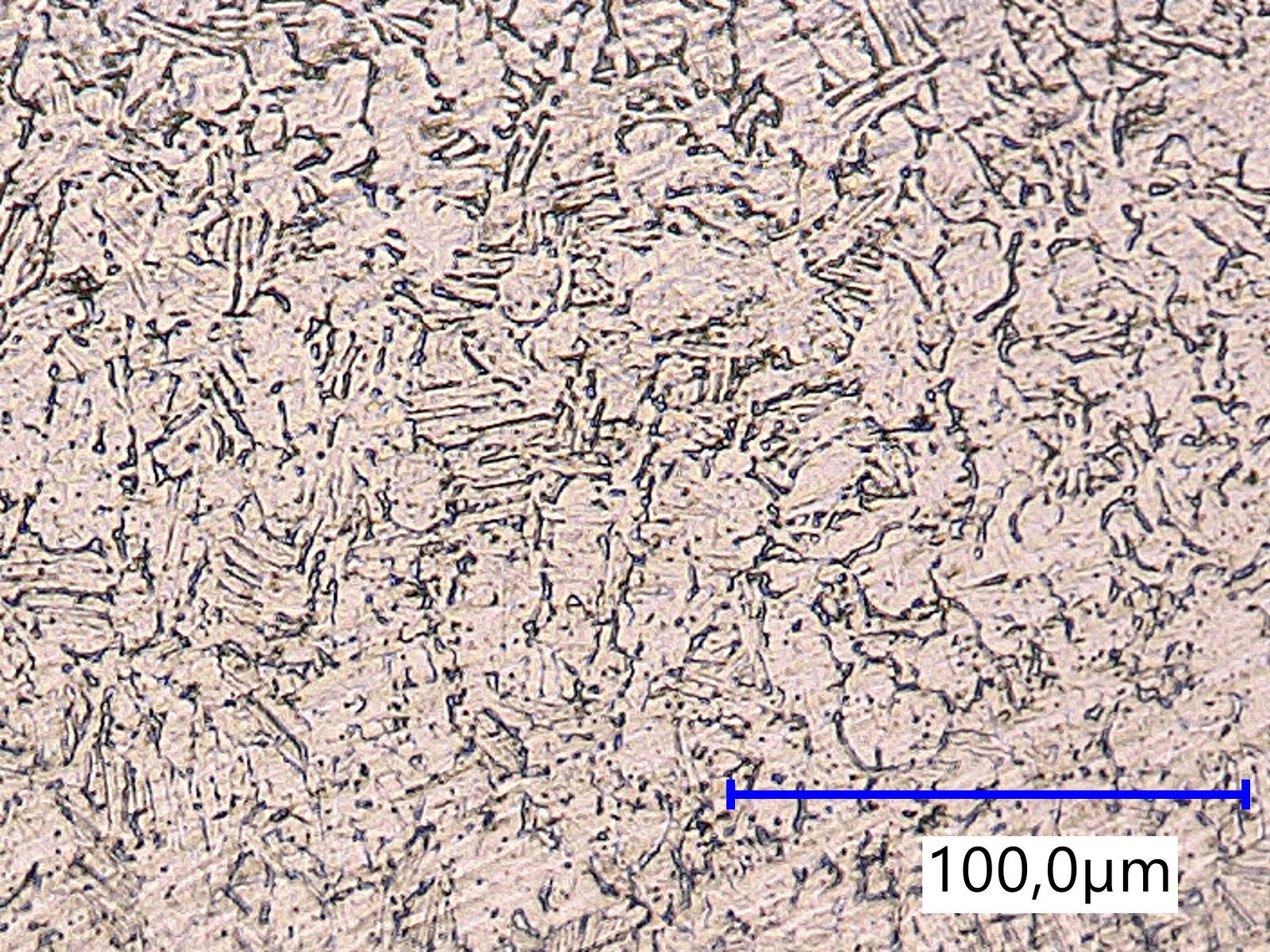}
	}
	\caption{Microstructure of \TitanPunkt: the top surface (left) and the side surface (right) show a uniform grain structure.}
	\label{Bild:Mikrostruktur_Ti6Al4V_Draufsucht_Seitenansicht}
\end{figure}

Etched samples are prepared for microstructural analyses of the top and side surface. The \Titan samples are etched with Kroll. The microstructure of the top and side surface are shown in Figure \ref{Bild:Mikrostruktur_Ti6Al4V_Draufsucht_Seitenansicht}. Due to the heat treatment both show a uniform microstructure without any salience.

\begin{figure}[b!]
	\center{
		\includegraphics[width=0.49\textwidth]{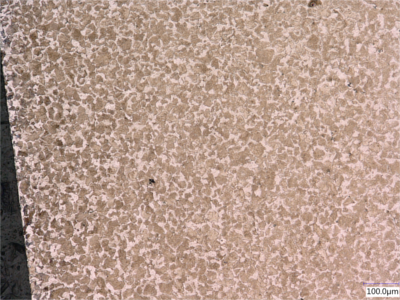}
		\includegraphics[width=0.49\textwidth]{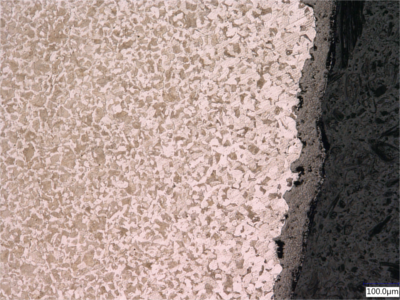}
	}
	\caption{Microstructure of Ck45 steel: top surface at the disk center (left) showing a uniform grain structure and at outer radius (right) showing decarburation and mill scale.}
	\label{Bild:Mikrostruktur_Ck45_Draufsicht_Zentrum_Auszenradius}
\end{figure}

\begin{figure}[b!]
	\center{ \includegraphics[width=0.8\textwidth]{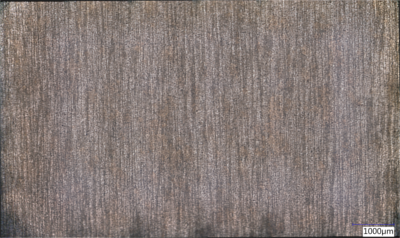}}
	\caption{Microstructure of Ck45 steel: side surface showing a columnar structure.}
	\label{Bild:Mikrostruktur_Ck45_Seitenansicht}
\end{figure}

The Ck45 is etched with Nital. Figure \ref{Bild:Mikrostruktur_Ck45_Draufsicht_Zentrum_Auszenradius} shows the top surface. A ferritic-perlitic microstructure is visible. Towards the outer surface, decarburations and mill scales can be seen. At the disk center the microstructure is uniform. The side surface microstructure revealed a columnar structure along the cylinder axis, see Figure \ref{Bild:Mikrostruktur_Ck45_Seitenansicht}. This is likely to have been induced by rolling during the manufacturing process.


\FloatBarrier

\subsubsection{EBSD}

\begin{figure}[h]
	\center{
		\includegraphics[width=0.49\textwidth]{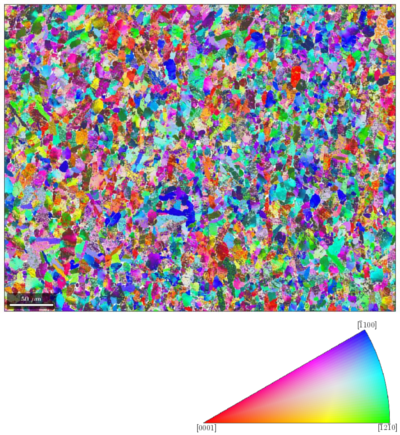}
		\includegraphics[width=0.49\textwidth]{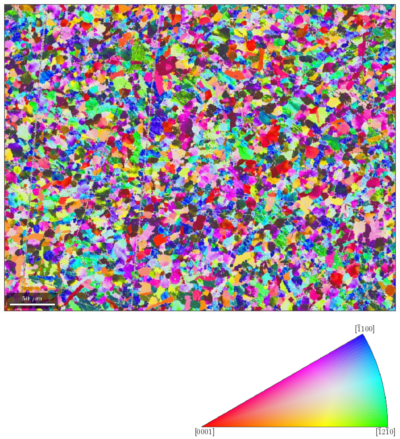}		
	}
	\caption{EBSD of \TitanPunkt: side (left) and top (right) surface with crystal orientations.}
	\label{Bild:EBSD_Ti6Al4V_Seitenansicht_Draufsicht}
\end{figure}

\begin{figure}[h]
	\center{
		\includegraphics[width=0.45\textwidth]{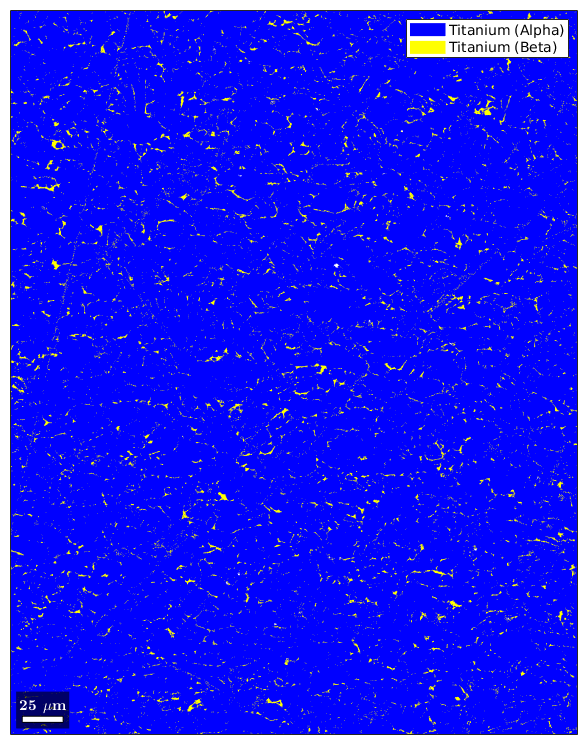}
		\includegraphics[width=0.45\textwidth]{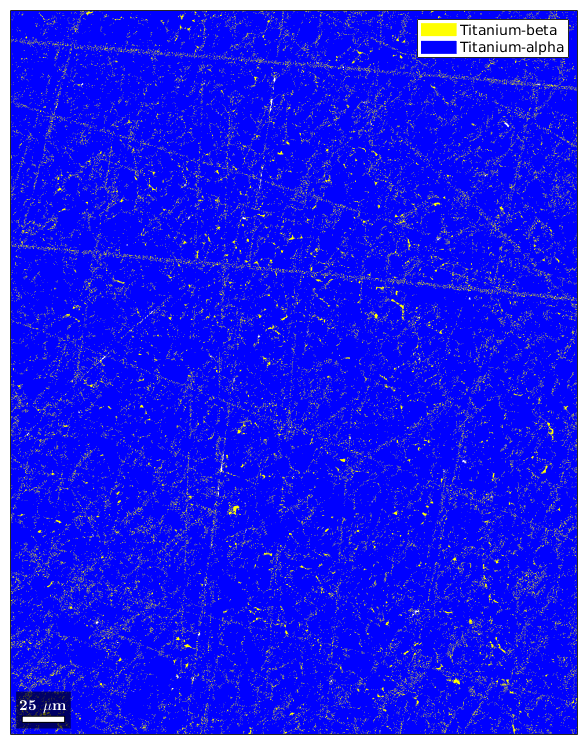}
	}
	\caption{EBSD of \TitanPunkt: $\alpha$ / $\beta$- phase distribution in the side (left) and top (right) surface.}
	\label{Bild:EBSD_Ti6Al4V_Seitenansicht_Draufsicht_AlphaBeta}
\end{figure}

\begin{figure}[h]
	\center{
		\includegraphics[width=0.75\textwidth]{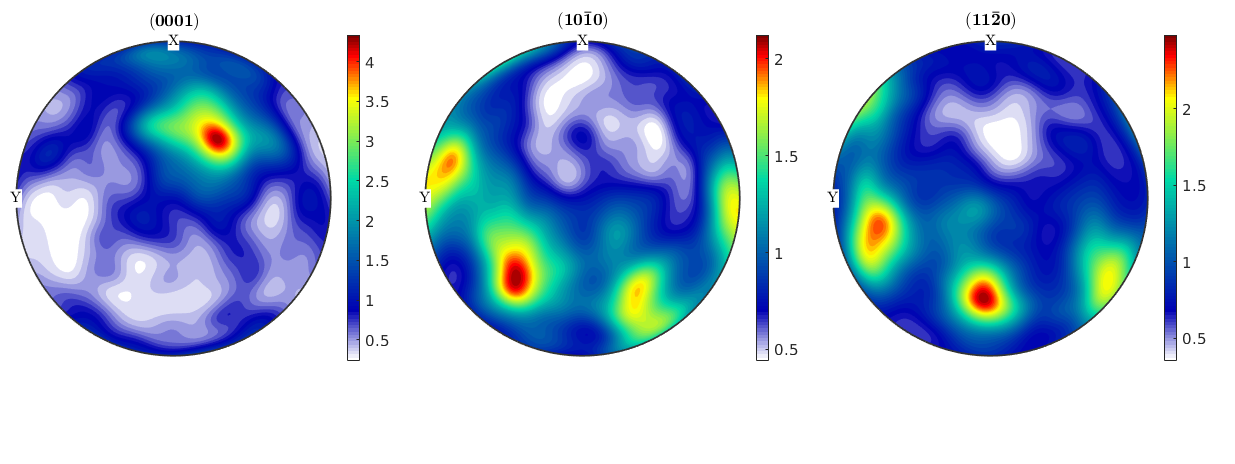}\\
		\includegraphics[width=0.75\textwidth]{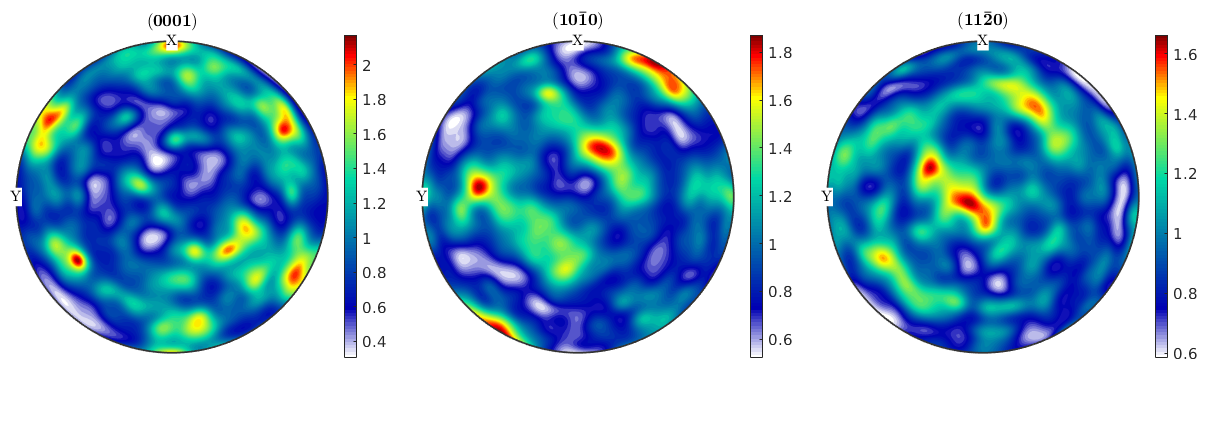}
	}
	\caption{EBSD of \TitanPunkt: side (top row) and top (bottom row) surface pole figures of the $\alpha$- phase.}
	\label{Bild:EBSD_Ti6Al4V_Seitenansicht_Draufsicht_Polfigur}
\end{figure}

\begin{figure}[h]
	\center{
		\includegraphics[width=0.49\textwidth]{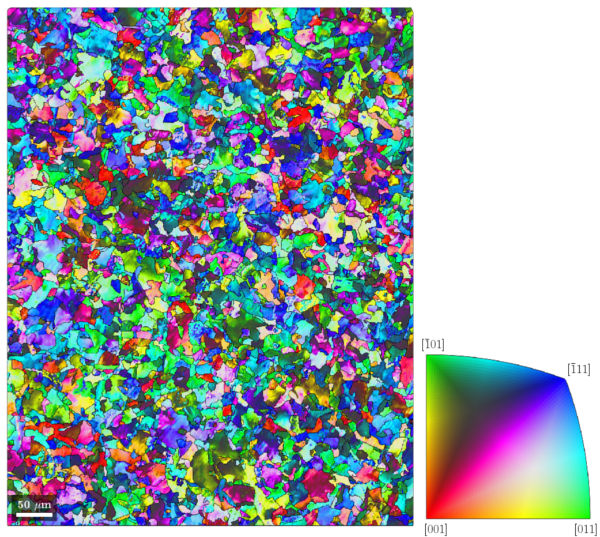}
		\includegraphics[width=0.49\textwidth]{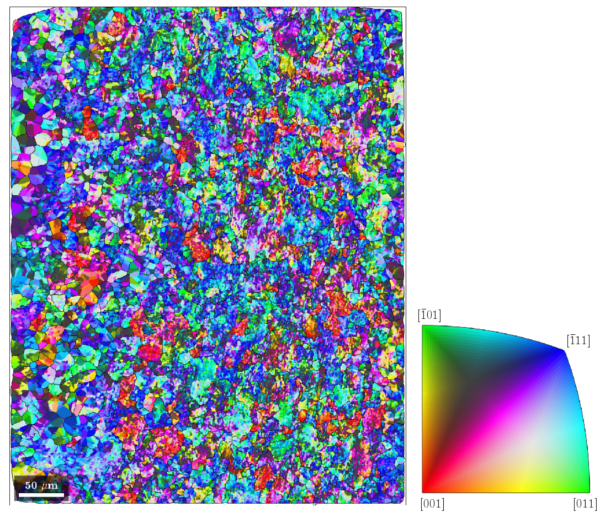}
	}
	\caption{EBSD of Ck45: top (left) and side (right) surface with crystal orientations.}
	\label{Bild:EBSD_Ck45_Draufsicht_Seitenansicht}
\end{figure}

\begin{figure}[h]
	\center{
		\includegraphics[width=0.75\textwidth]{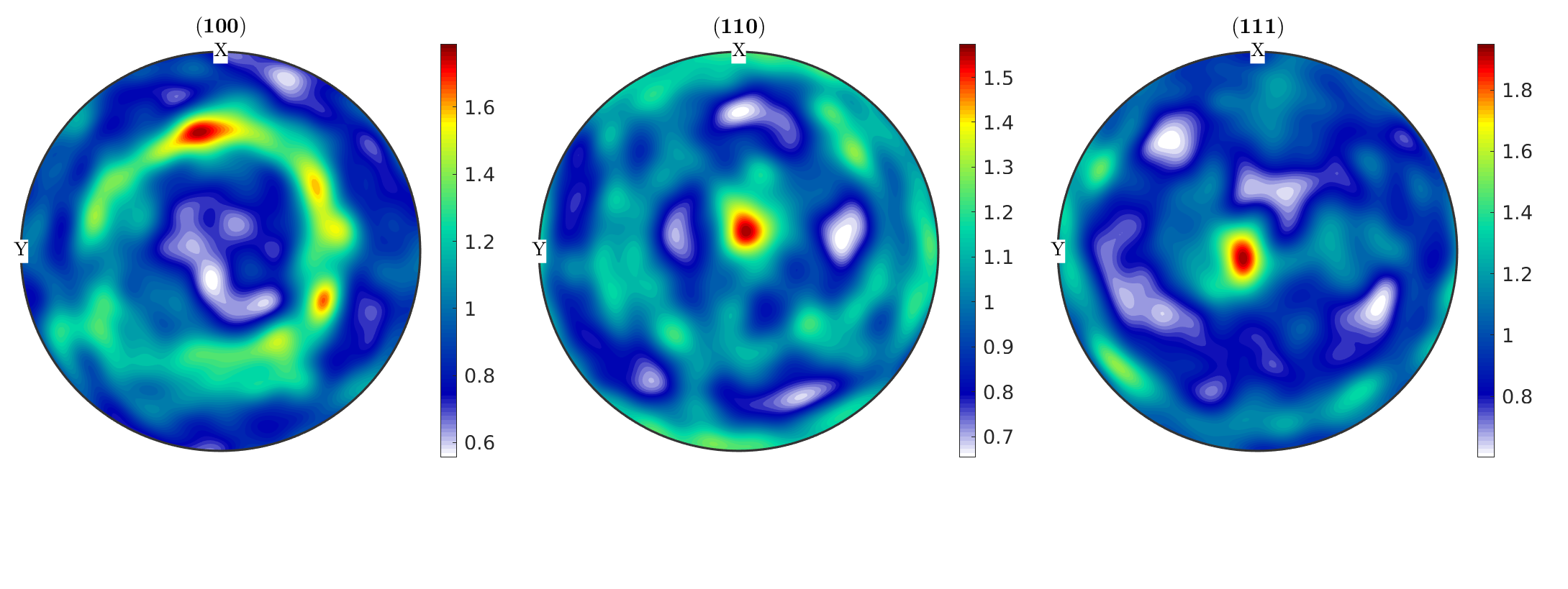}
		\includegraphics[width=0.75\textwidth]{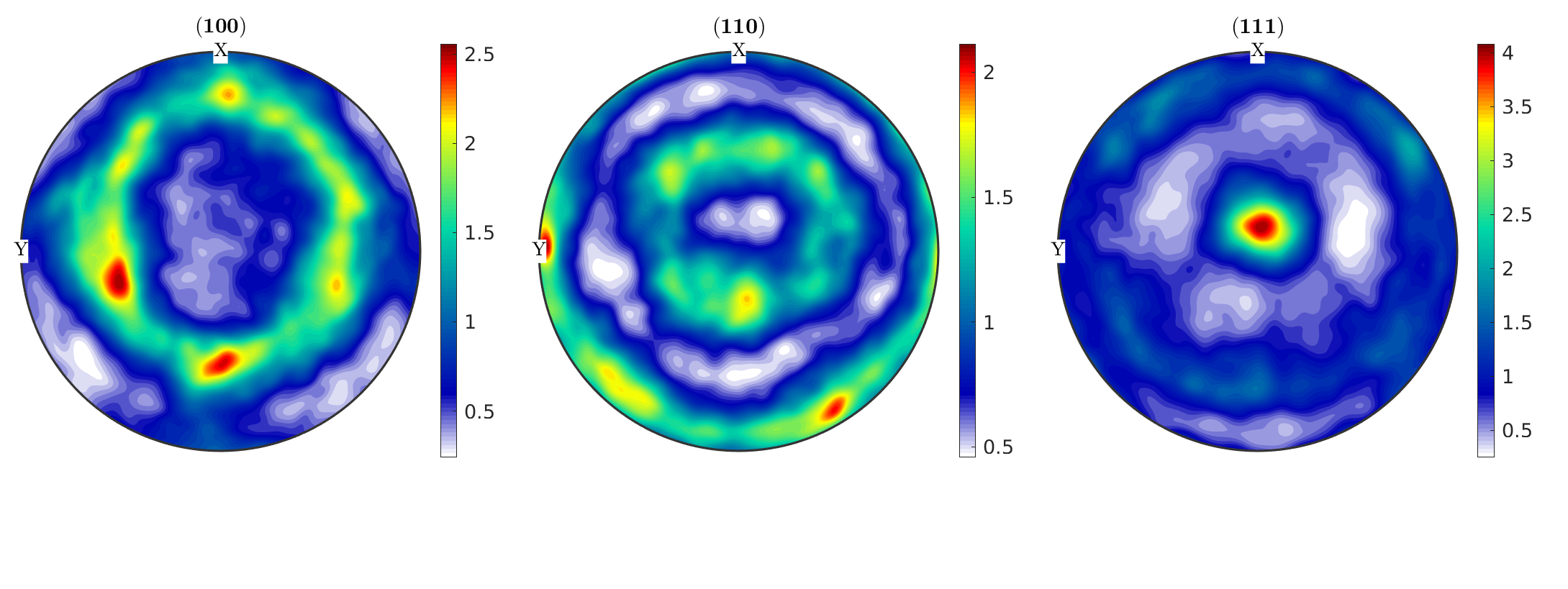}
	}
	\caption{EBSD of Ck45: top (top row) and side (bottom row) surface pole figure.}
	\label{Bild:EBSD_Ck45_Draufsicht_Seitenansicht_Polfigur}
\end{figure}

The crystallographic orientations in the grains and the textures are measured with EBSD for the top and side surfaces of the Ck45 and \Titan material. The measured areas of the samples are given in table \ref{Tab:EBSD_Meszbereich}. The EBSD-analysis of \Titan reveals a slightly different ratio of $\alpha$- and $\beta$-phase for top and side surface. The average grain diameter and average aspect ratios of the grains are similar for the side and top surface. The results, together with grain sizes and aspect ratios, are given in table \ref{Tab:EBSD_Ergebnisse_Ti6Al4V_Ck45} for \Titan and Ck45. The grain orientations of \Titan are shown for the side and top surface in Figure \ref{Bild:EBSD_Ti6Al4V_Seitenansicht_Draufsicht}, the corresponding distributions of $\alpha$ and $\beta$-phases in Figure \ref{Bild:EBSD_Ti6Al4V_Seitenansicht_Draufsicht_AlphaBeta}. The pole figures of the $\alpha$-phases are given with Figures \ref{Bild:EBSD_Ti6Al4V_Seitenansicht_Draufsicht_Polfigur} for side and top surface, respectively. The pole figures reveal a slightly stronger texture in the side surface than in the top surface. The grain orientations of Ck45 are shown for the top and side surface in Figure \ref{Bild:EBSD_Ck45_Draufsicht_Seitenansicht} and the pole figures in Figure \ref{Bild:EBSD_Ck45_Draufsicht_Seitenansicht_Polfigur}. Similar to \TitanPunkt, the side surface shows a stronger texture than the top surface which may have been induced by the rolling process during manufacturing.


\begin{table}[h]
	\scriptsize
	\begin{center}
		\begin{tabular}{ c | c | c | c | c }
			Specimen & Surface & Sample size $[\mu m^2]$ & Figures \\
			\hline
			\Titan	& Side surface	& 443 x 347 & \multirow{2}{*}{\ref{Bild:EBSD_Ti6Al4V_Seitenansicht_Draufsicht}, \ref{Bild:EBSD_Ti6Al4V_Seitenansicht_Draufsicht_AlphaBeta}, \ref{Bild:EBSD_Ti6Al4V_Seitenansicht_Draufsicht_Polfigur}}\\
			\Titan	& Top surface	& 331 x 428 & \\
			\hline
			Ck45	& Side surface		& 545 x 428 & \multirow{2}{*}{\ref{Bild:EBSD_Ck45_Draufsicht_Seitenansicht}, \ref{Bild:EBSD_Ck45_Draufsicht_Seitenansicht_Polfigur}}\\
			Ck45	& Top surface		& 709 x 556 & \\
		\end{tabular}
		\caption{EBSD measurement sizes of the four specimen.}
		\label{Tab:EBSD_Meszbereich}
	\end{center}
\end{table}

%

\begin{table}[h]
	\scriptsize
	\begin{center}
		\begin{tabular}{ c |  c | c | c | c | c | c | c | c}
			Material & Surface & Avg. equiv. grain & Std.dev. & Avg. grain & Std. dev. grain & $\alpha$-phase & $\beta$-phase & Comment\\
			& & diameter $D_g [\mu m]$ & $\mu_{D_g} [\mu m]$ & aspect ratio & aspect ratio & vol. $\%$ & vol.$\%$ \\
			\hline
			\Titan & Side surface	& 9.9 & 4.2 & 1.8 & 0.6 & 96.7 & 3.3 \\
			\Titan & Top surface	& 9.1 & 3.2 & 1.7 & 0.5 & 94.8 & 5.2 \\
			Ck45 & Side surface	& 17.8 & 13.5 & 1.8 & 0.5 & - & - & noisy raw data\\
			Ck45 & Top surface		& 14.3 & 9.3 & 1.8 & 0.6 & - & - &\\
		\end{tabular}
		\caption{\Titan and Ck45: EBSD measurement results for side and top surface.}
		\label{Tab:EBSD_Ergebnisse_Ti6Al4V_Ck45}
	\end{center}
\end{table}

%

%
%

\FloatBarrier

%
%
%
%


\subsection{Tensile Tests}
\label{Kap:ZugversucheIVP}

\begin{figure}[h]
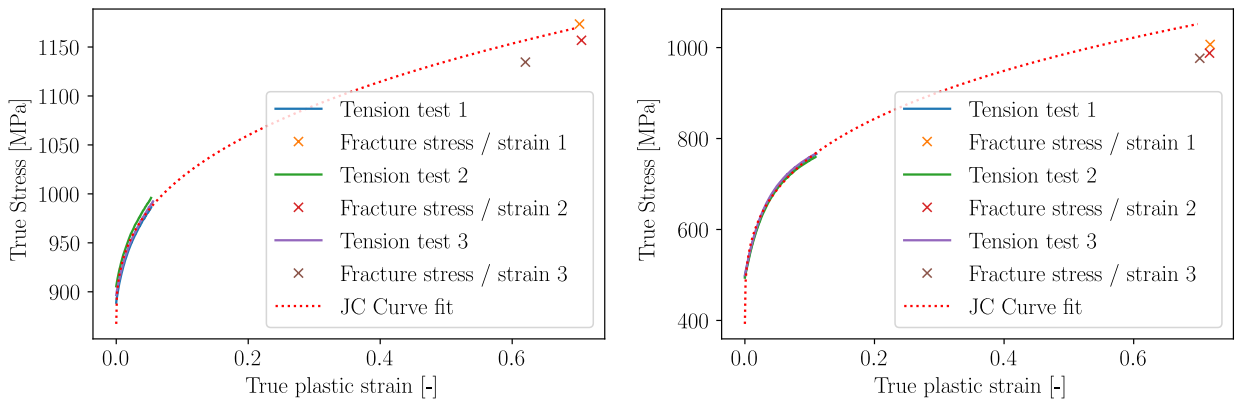

	\begin{center}
		\includesvg[width=0.49\textwidth]{Bilder/Ti6Al4V_JC_ABn_Anpassung_Veroeffentlichung_0_002s}
		\includesvg[width=0.49\textwidth]{Bilder/Ck45_JC_ABn_Anpassung_Veroeffentlichung_0_002s}
	\end{center}
	\caption{Fit of the Johnson-Cook parameters A, B and n for quasi-static conditions ($\dot \varepsilon_{pl}\!=\!0.002s^{-1}$) for \Titan (left) and Ck45 (right).}
	\label{Bild:Ti6Al4V_ABn_JC_Anpassung}
\end{figure}

\begin{figure}[h]
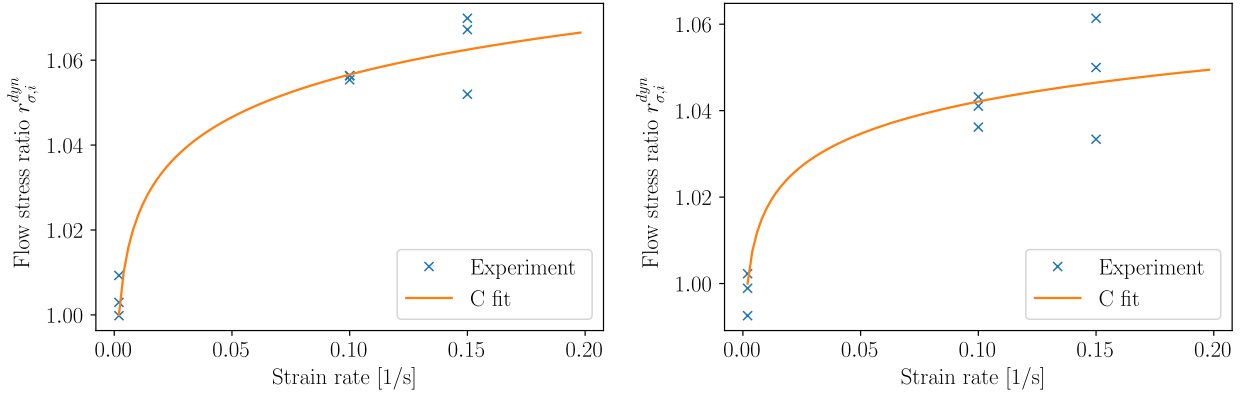

	\begin{center}
		\includesvg[width=0.49\textwidth]{Bilder/Ti6Al4V_JC_C_Anpassung}
		\includesvg[width=0.49\textwidth]{Bilder/Ck45_JC_C_Anpassung}
	\end{center}
	\caption{Fit of the Johnson-Cook strain rate sensitivity parameter C for \Titan (left) and Ck45 (right).}
	\label{Bild:Ti6Al4V_C_JC_Anpassung}
\end{figure}

In the following, tensile test results on this batch of materials are summarized from \cite{Klippel2021}. The tensile tests are performed for both materials, Ck45 and \TitanPunkt, at room temperature for three different strain rates ($\dot \varepsilon_{pl}=\SI{0.002}{\per \second},\SI{0.1}{\per \second},\SI{0.15}{\per \second}$) and each repeated three times. The tensile test specimen are produced from the inner core of the cylinders according to DIN50125 \cite{DIN50125} with form B and the dimensions B8x40. The tensile tests are used to fit material parameters for the work hardening and strain rate sensitivity part of the Johnson-Cook flow stress model \cite{JohnsonCook1983}:

\begin{equation}
	\label{Glg:JohnsonCook}
	\sigma_y = \underbrace{\left( A+B \cdot (\varepsilon_{pl})^n\right)}_{\text{work hardening}} \hspace{3mm} \underbrace{\left( 1+C \cdot ln\left(\frac{\dot\varepsilon_{pl}}{\dot\varepsilon^0_{pl}}\right) \right)}_{\text{strain rate sensitivity}} \hspace{3mm} \underbrace{\left(1- \left(\frac{T-T_{ref}}{T_f-T_{ref}} \right)^m \right)}_{\text{thermal softening}}
\end{equation}

where A, B, C, m and n are material dependent parameters, $\varepsilon_{pl}$ is the equivalent plastic strain, $\dot \varepsilon_{pl}$ the plastic strain rate, $T$ the temperature, $T_f$ the melting temperature and $T_{ref}$ the reference temperature of the material tests. Tests at temperatures higher than room temperature are not carried out, therefore the parameter $m$ leaves undetermined. The parameters $A$, $B$, $C$ and $n$ are determined with the procedure in \cite{Klippel2020}. The strain rate sensitivity parameter C is valid only at very low strain rates, as the tests are conducted in the strain rate range from $\SI{0.002}{\per \second}$ to $\SI{0.15}{\per \second}$. The obtained Johnson-Cook parameters are given in table \ref{Tab:Ti6Al4VCk45_ZugversucheIVP_JC_ABCn} and the
approximated flow curves are given in Figure \ref{Bild:Ti6Al4V_ABn_JC_Anpassung} for \Titan and Ck45. The curve fits of the strain rate sensitivities are shown in Figure \ref{Bild:Ti6Al4V_C_JC_Anpassung}.

\begin{table}[h]
	\small
	\begin{center}
		\begin{tabular}{c | c | c | c | c | c}
			Material & A [MPa] & B [MPa] & n [-] & $\dot{\varepsilon}_{pl}^{ref} [s]$ & C [-]\\
			\hline
			\Titan & 867 & 344 & 0.361 & 0.002 & 0.0145\\
			Ck45 & 392 & 735 & 0.304 & 0.002 & 0.0108
		\end{tabular}
		\caption{\Titan and Ck45: fit of the work hardening parameters A, B and n and the strain rate sensitivity C of the JC flow stress model \eqref{Glg:JohnsonCook}.}
		\label{Tab:Ti6Al4VCk45_ZugversucheIVP_JC_ABCn}
	\end{center}
\end{table}

\FloatBarrier

\subsection{Cutting Tests}
\label{Kap:Zerspanungsversuche}

\begin{figure}[h]
	\center{
		\includegraphics[width=\textwidth]{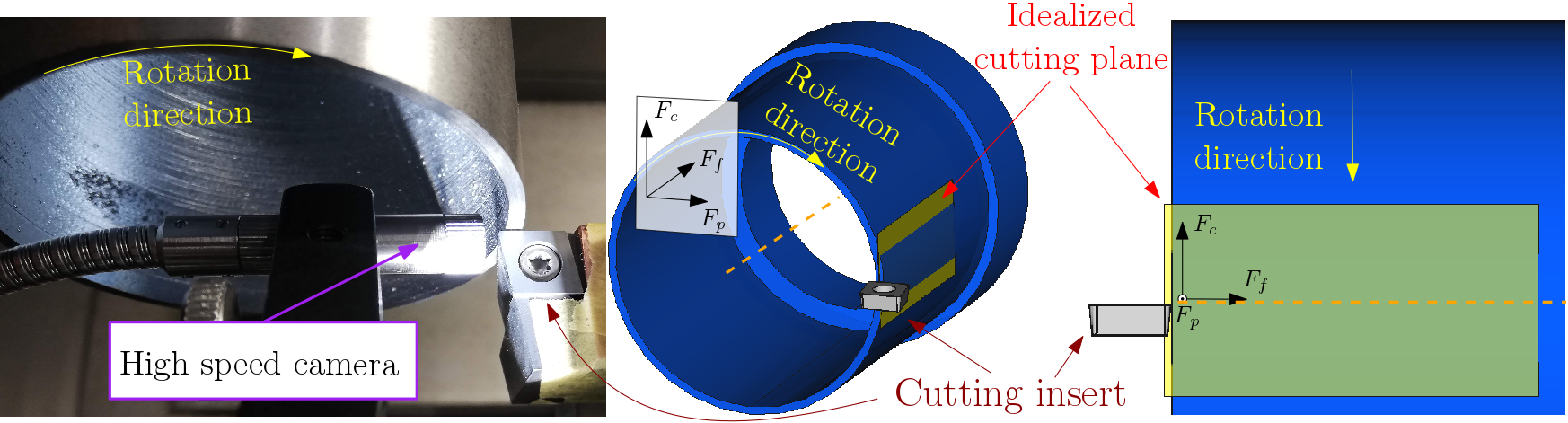}
	}
	\caption{Quasi-orthogonal cutting setup: cylinder with cutting insert (left and middle), cutting plane in 2D (right).}
	\label{Bild:Orthogonalschnitt_Entwicklung}
\end{figure}

For the machining test, thin-walled cylinders with a diameter of $D \approx \SI{72}{\milli\meter}$ and a wall thickness of $d \approx \SI{2}{\milli\meter}$ are turned longitudinally on a Schaublin 42L CNC lathe. This emulates an orthogonal cut closely. The cylinders are turned inside and outside from the raw workpieces in the same clamping setup. The machining of the outer surfaces removes the scaling and the decarburated zone of the Ck45. The setup with a cylinder in place is shown in Figure \ref{Bild:Orthogonalschnitt_Entwicklung} together with the experimental setup for the orthogonal cutting tests. Forces are measured using a Kistler 9121A5 dynamometer in combination with a Kistler 5019A charge amplifier and digitalized with a NI USB-6211 data acquisition system. The signal is low-pass filtered with a cut-off frequency of 30Hz before being sampled with 1.1 kHz. Offset and drift in the measured force signal $\tilde{F}_{meas}$ are corrected after the measurement since small drifts $\tilde{F}_{drift}$ in the signal can occur and overlay the process forces $\tilde{F}_{proc}$:

\begin{equation}
	\label{Glg:Kraftsignal_Korrektur}
	\tilde{F}_{proc} = \tilde{F}_{meas} - \tilde{F}_{drift}
\end{equation}

\begin{figure}[h]
	\center{\includegraphics[width=0.6\textwidth]{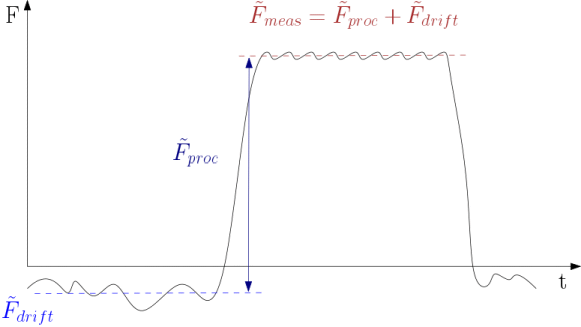}}
	\caption{Force signal overlayed with a drift.}
	\label{Bild:Kraftkorrektur}
\end{figure}

The drift $\tilde{F}_{drift}$ of the force is evaluated by positioning the cutter before the cut with a distance of one mm away from the cylinder to cut. With the desired feed rate of the experiment the cutter approached the workpiece and during this time the idle-forces $\tilde{F}_{drift} = \tilde{F}_{idle}$ are recorded and then used to correct to force signals with \eqref{Glg:Kraftsignal_Korrektur}. A schematics of this correction is shown in Figure \ref{Bild:Kraftkorrektur}.

\FloatBarrier

High speed camera recordings are conducted for the first cutting tests with a Phantom V12.1 camera using an endoscope, see the cutting setup in Figure \ref{Bild:Orthogonalschnitt_Entwicklung}. References to these records are linked in the result tables \ref{Tab:TestResults_Ti6Al4Vkpl} and \ref{Tab:TestResults_Ck45kpl}. In order to simplify numerical modelling, the experiments are performed as dry cuts without the use of a lubricant. To date no viable approach for considering coolant and lubricant in numerical modeling exists, and the use of dry sections reduces the number of unknown parameters. Each combination of feed $f$ and cutting speed $v_c$ is usually tested three times to ensure results quality and repeatability. For each test an unused cutting edge is used and test durations are kept short so that wear stays insignificant.


\begin{figure}[h]
	\begin{center}
		\includegraphics[width=0.3\textwidth]{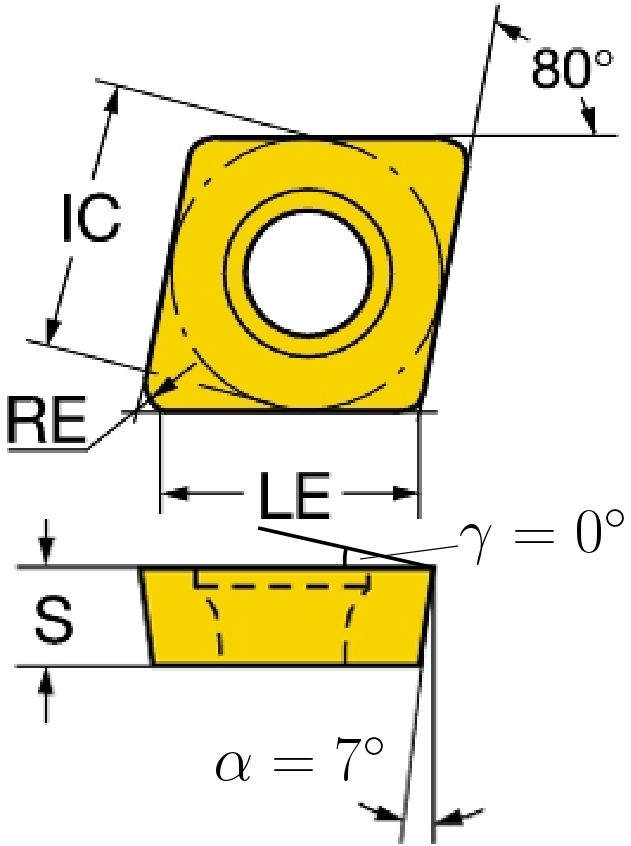}
		\includegraphics[width=0.4\textwidth]{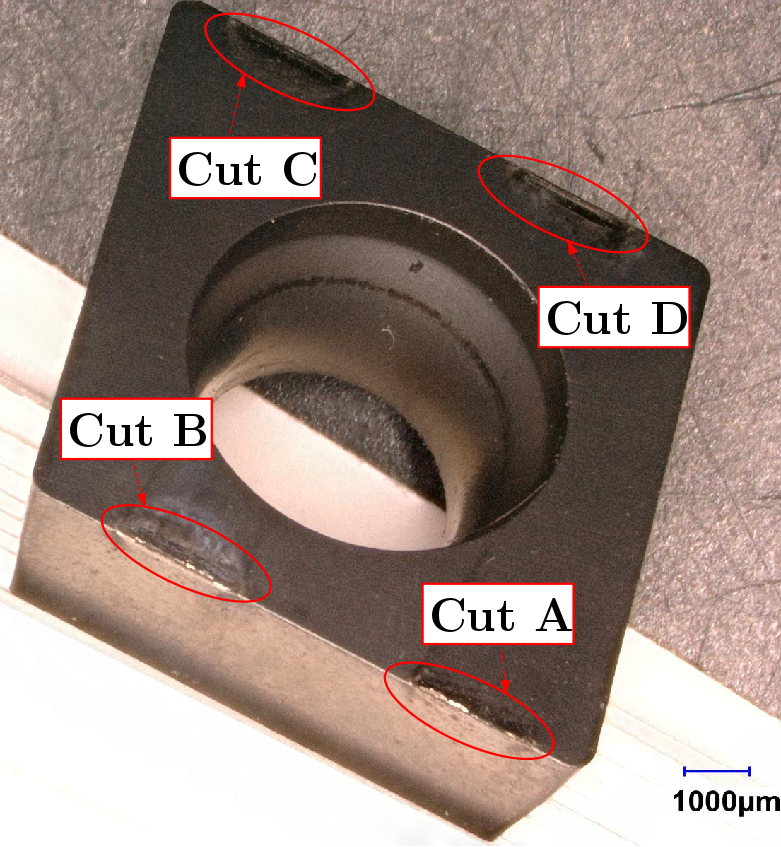}
	\end{center}
	\caption{Cutting insert geometry (left) from \cite{Sandvik} and four cut positions A-D (right) for cutting experiments, from \cite{Klippel2021}.}
	\label{Bild:Wendeschneidplatte_Zerspanungsversuche}
\end{figure}

Uncoated turning inserts \textit{Sandvik Coromant} CCMW 09T304 H13A (ISO) are used for the cutting experiments.
The main geometrical data of the inserts are given in table \ref{Tab:WSP_Geometrie} and pictures of the turning insert geometry are provided in Figure \ref{Bild:Wendeschneidplatte_Zerspanungsversuche}.

\begin{table}[ht]
	\footnotesize
	\centering
	\begin{tabular}{ c | c | c | c | c | c | c}
		Edge & Cutting edge & Inscribed & Cutting edge & Clearance & Rake & Cutting edge\\
		radius RE & height S & circle IC & length LE & angle $\alpha$ & angle $\gamma$ & radius\\
		\hline
		$0.397mm$ & $3.969mm$ & $9.525mm$ & $9.272mm$ & $7^\circ$ & $0^\circ$ & see chapter \ref{Kap:OptischeVermessungWSP}
	\end{tabular}
	\caption{Main geometry data of \textit{Sandvik Coromant} CCMW 09T304 H13A (ISO) inserts.}
	\label{Tab:WSP_Geometrie}
\end{table}

The cutting edge radii along the cutting edge length LE result from the sintering process and exhibit a certain scatter. Since these radii have a significant impact on the process forces, they are optically measured, see details in chapter \ref{Kap:OptischeVermessungWSP}. An Applitec SCACL-2020X-09 tool holder (ISO-2216) is used, which has an entering angle $\xi = \SI{90}{\degree}$ and an inclination angle $\kappa = \SI{0}{\degree}$. Each insert is used for four cutting experiments (two cuts per side of the insert) where after every experiment another position A-D was used on the insert. These four cut positions A-D are shown in Figure \ref{Bild:Wendeschneidplatte_Zerspanungsversuche}.


\begin{figure}[h]
	\begin{center}
		\includegraphics[width=0.8\textwidth]{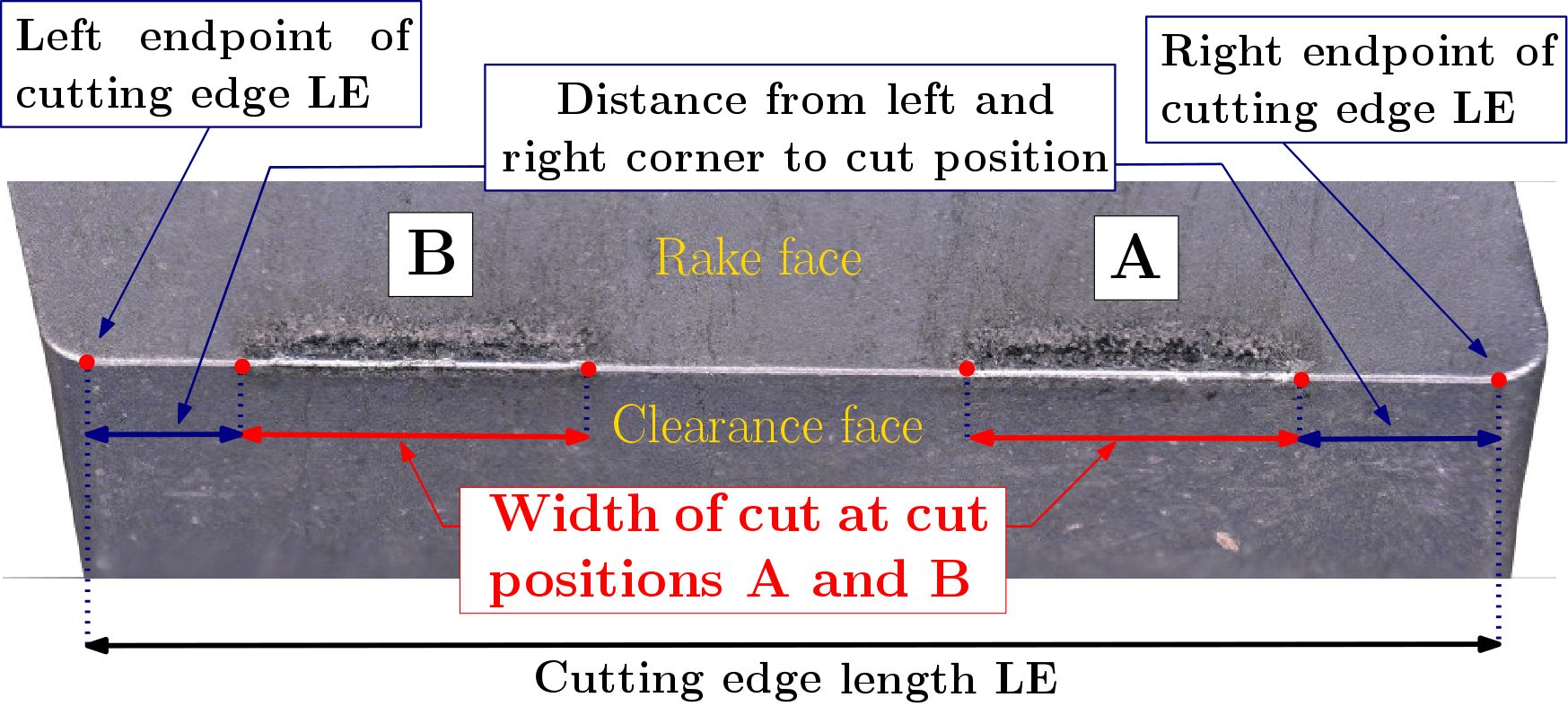}
	\end{center}
	\caption{Cutting insert with cut positions A and B showing start and endpoints of the cutting edge, shortest distances from the corners to the beginning of the cutting zones and width of cuts.}
	\label{Bild:Wendeschneidplatte_Meszpunkte}
\end{figure}

%
%
%
%
\subsubsection{Determination of Cutting Edge Radii}
\label{Kap:OptischeVermessungWSP}

\begin{figure}[htp]
	\center{
		\includesvg[width=0.8\textwidth]{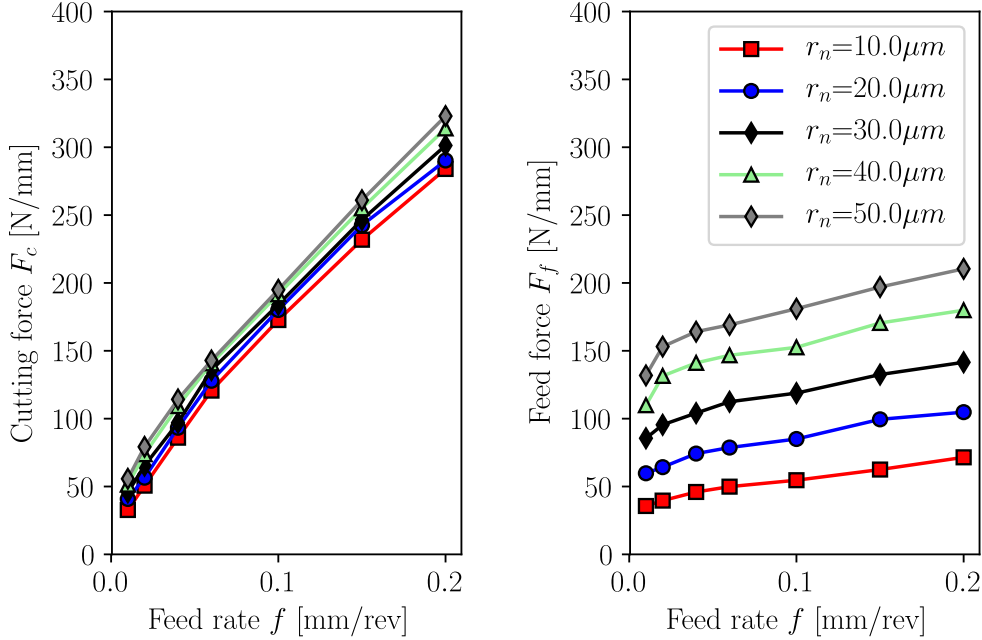}
	}
	\caption{Influence of cutting edge radius $r_n$ and feed $f$ on the cutting and feed forces for turning of \TitanPunkt. The forces are standardised to a cut width of $b=\SI{1}{\milli\meter}$, the cut speed was $v_c=\SI{70}{\meter/\minute}$, the rake angle $\gamma=10^\circ$ and the clearance angle $\alpha=8^\circ$ according to measurements from \cite{Wyen2010}.}
	\label{Bild:Wyen_Schneidkantenradius_Einflusz_2010}
\end{figure}

The shape of the cutting insert, especially the cutting edge radius, have a major influence on the process forces of the turning operation. Albrecht \cite{Albrecht1960} showed for steel that with increasing cutting edge radius the feed force $F_f$ increases strongly and the cutting force $F_c$ increases moderately. The effect becomes more pronounced with higher feeds. Wyen \cite{Wyen2010} conducted a similar investigation for \Titan revealing comparable influences on the process forces, see Figure \ref{Bild:Wyen_Schneidkantenradius_Einflusz_2010}.

%

The inserts in this investigation have an unspecified cutting edge radius and therefore all cutting edges (2 x 149 inserts) of the cutting inserts are optically measured with an Alicona InfiniteFocusG4 microscope prior to the cut tests in an unused condition. With the focus variation principle \cite{DIN25178,Klocke2018}, 3D data of the scanned surfaces are generated, where a 20x magnification with $\SI{0,5}{\micro\meter}$ vertical and $\SI{7}{\micro\meter}$ lateral resolution is used to scan along the complete cutting edge length (LE). Initially a scan depth of $\SI{60}{\micro\meter}$ was used but since in some positions cutting edge radii of about $\SI{50}{\micro\meter}$ appeared, it was then later increased to $\SI{400}{\micro\meter}$. Before the cutting edge radii determination the scanned 3D data is reworked. In a first step single peaks (measurement artefacts) are removed. This is followed by the determination of the left and right endpoints of the cutting edge (LE) between the corner radii for which height profile gradients are evaluated. In between these two points the cutting edge radii are analyzed and they serve as reference points for the exact determination of the cut position after the experiments. 

\begin{figure}[h]
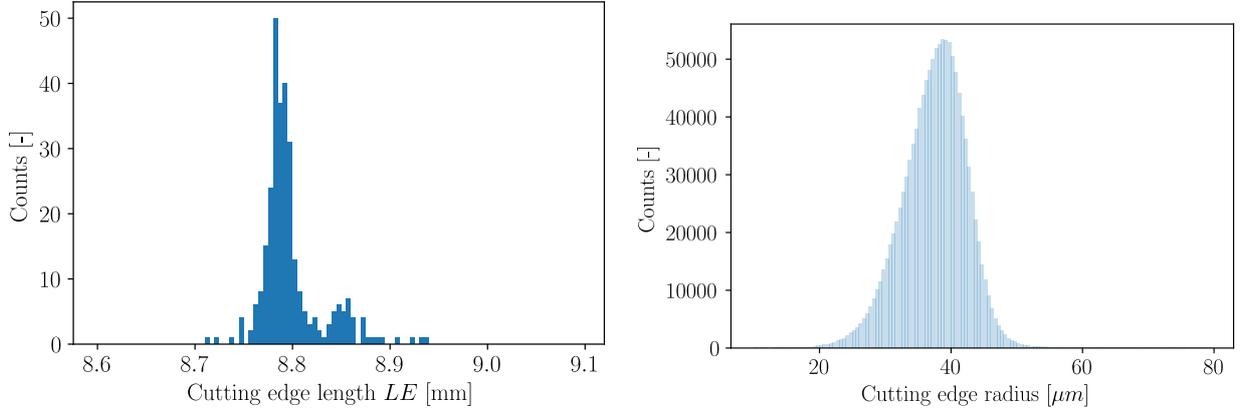

	\begin{center}
		\includesvg[width=0.49\textwidth]{Bilder/HistogrammSchneidkantenbreiten_149WSP_Sandvik_CCMW09T304H13A_AliconaMessung}
		\includesvg[width=0.49\textwidth]{Bilder/HistogrammSchneidkantenradien_149WSP_Sandvik_CCMW09T304H13A_AliconaMessung}
	\end{center}
	\caption{Histogram of cutting edge lengths $LE$ (left) and cutting edge radii (right) over all cutting inserts.}
	\label{Bild:Schneidkantenradiusmessung_WyenVerfahren}
\end{figure}

\begin{figure}[h]
	\begin{center}
		\includegraphics[width=0.8\textwidth]{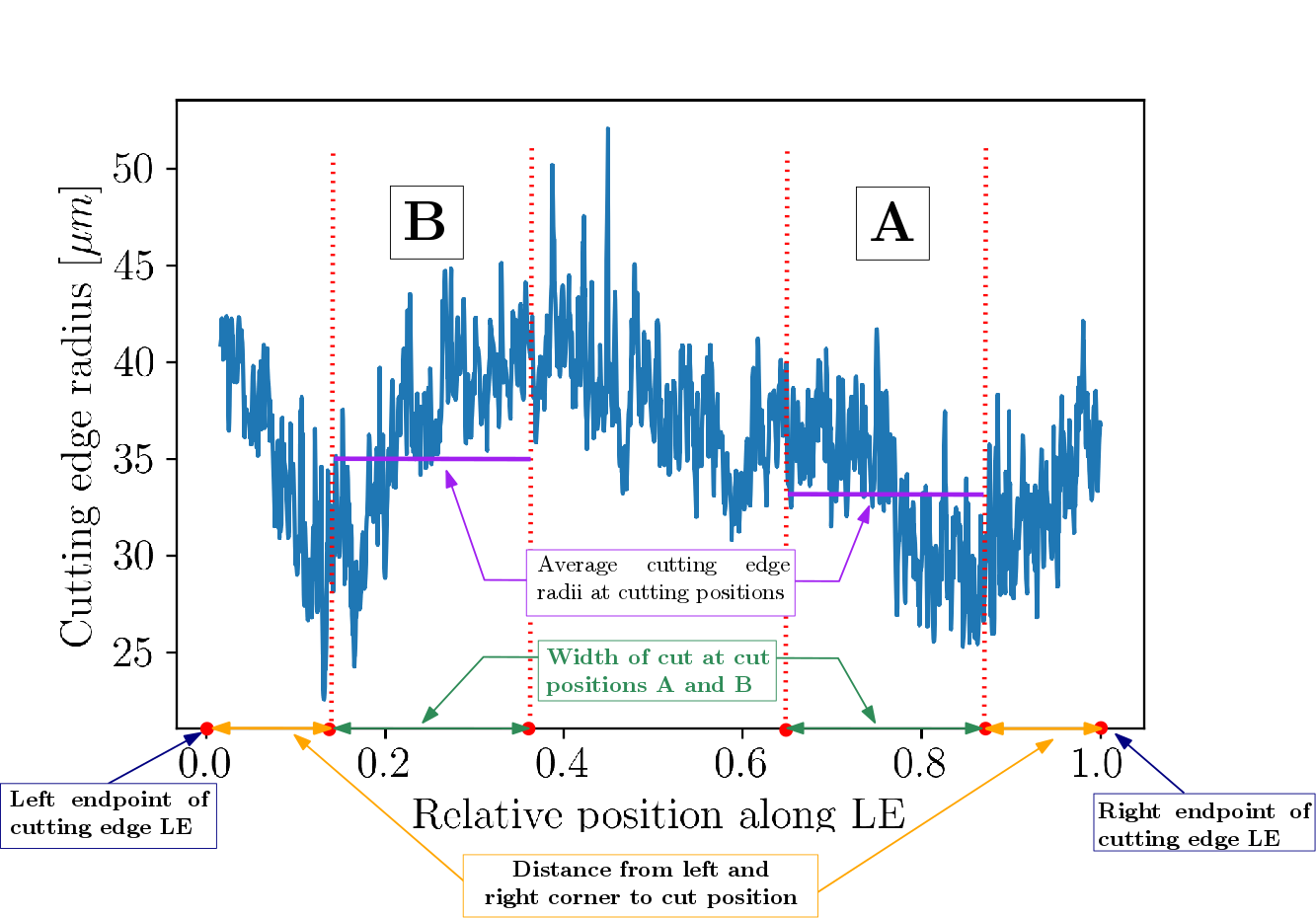}
	\end{center}
	\caption{Example of cutting edge radii variation along a single cutting edge, with averaged cutting edge radii at cut positions A and B.}
	\label{Bild:Schneidkantenradiusmessung_WSP116AB_ueberLE}
\end{figure}

The determination of the cutting edge radii follows the procedures outlined in \cite{Meier2019,Wyen2012} in order to ensure reproducibility of the results. A total of about 1'275'000 cutting edge radii ($>4000$ radii per cutting edge) are extracted from the scans. Over all 149 cutting inserts, the cutting edge radius along LE varies between $\approx \SI{20}{\micro\meter}$ and $\approx \SI{50}{\micro\meter}$. The variation of the cutting edge radius along one cutting edge length $LE$ can be significant, as for instance shown in Figure \ref{Bild:Schneidkantenradiusmessung_WSP116AB_ueberLE}, where it varies between $r_n \approx \SI{25}{\micro\meter}$ and $r_n \approx \SI{50}{\micro\meter}$. The mean cutting edge radius is $\SI{37,5}{\micro\meter}$ with a standard deviation $\SI{4,9}{\micro\meter}$. The scatter of the radius is the reason, why the exact cutting edge radius was determined after the experiments, when the location of the cut on the cutting edge is known for every insert and its four cut positions A-D. The insert's edge radii serve as reference points to determine the start point of the cut positions. From this start point the cutting edge radii are averaged along the cut width of $\approx \SI{2}{\milli\meter}$. Its averaged values along the cut width, including the standard deviation, are given for each cutting experiment in tables \ref{Tab:TestResults_Ti6Al4Vkpl} and \ref{Tab:TestResults_Ck45kpl}. Histogram plots of the cutting edge length $LE$ and the cutting edge radii are given in Figure \ref{Bild:Schneidkantenradiusmessung_WyenVerfahren}. The length of the cutting edge $LE$ varies between $\approx \SI{8,7}{\milli\meter}$ and $\approx \SI{8,95}{\milli\meter}$ which is inline with the tolerance class of the selected inserts.


\FloatBarrier

\subsubsection{Chip Thickness Measurements}
\label{Kap:Spandickenmessung}

\begin{figure}[h]
	\centering
	\includegraphics[width=0.5\textwidth]{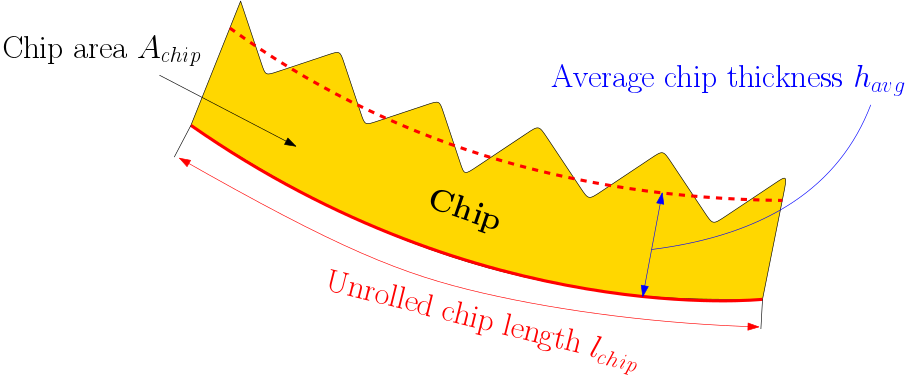}
	\caption{Measurement of average chip thicknesses $h_{avg}$ by the chip area $A_{chip}$ and unrolled chip length $l_{chip}$.}
	\label{Bild:Spanvermessung}
\end{figure}

At least one chip from the three repetitions of each set of feed $f$ and cut speed $v_c$ is embedded, ground (120, 240, 500, 1000, 2500, 4000) and polished ($6\mu m$, $3\mu m$, $1\mu m$). After polishing, the chips are etched with Kroll (\TitanPunkt) and Nital (Ck45). The geometry and microstructure are then analysed with a Keyence VHX-5000 microscope. The main dimensions which are measured are the chip area $A_{chip}$ and the unrolled chip length $l_{chip}$, see Figure \ref{Bild:Spanvermessung}, from which the average chip thickness $h_{avg}$ is computed as:

\begin{equation}
	\label{Glg:Spandickenmessung}
	h_{avg} = \frac{A_{chip}}{l_{chip}}
\end{equation}

\subsubsection{Test Plan}

\begin{figure}[h]
	\begin{center}
		\includesvg[width=0.6\textwidth]{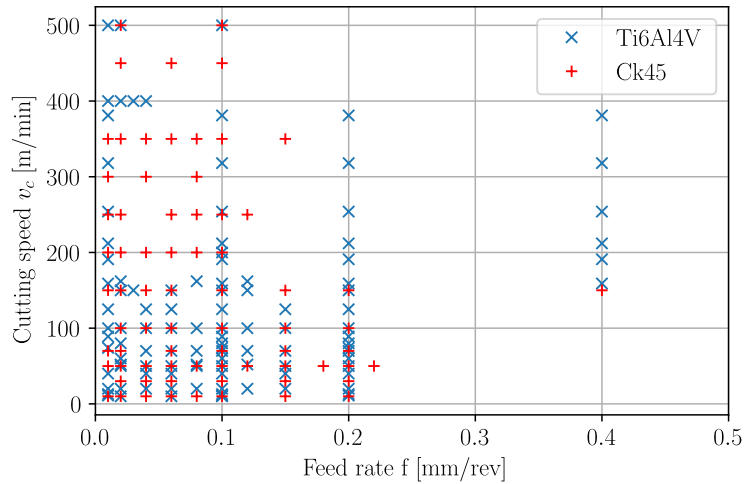}
	\end{center}
	\caption{Process parameter combinations used in the orthogonal cutting experiments.}
	\label{Bild:Zerspanungsversuche_Versuchsplan}
\end{figure}

Orthogonal cutting tests are conducted 
for a large range of feed and cutting speeds.
Each experimental set is repeated three times to ensure the reproducibility of the results. In total 520 cutting tests are conducted, of which 288 are for \Titan and 232 for Ck45, respectively. The parameter ranges are given in table \ref{Tab:TestMatrix} and all combinations of feed $f$ and cutting speeds $v_c$ which are used in the cutting experiments of \Titan and Ck45 are displayed in Figure \ref{Bild:Zerspanungsversuche_Versuchsplan}.

\begin{table}[h]
	\centering
	\begin{tabular}{ c | c | c }
		Material & Cutting speed $v_c [m\!/\!min]$ & Feed rate $f$ [mm/rev]\\
		\hline
		\Titan & 10...500 & 0.01...0.4\\
		Ck45 & 10...500 & 0.01...0.4\\
	\end{tabular} 
	\caption{Test matrix of the cutting tests}
	\label{Tab:TestMatrix}
\end{table}

\FloatBarrier

\section{Results}
\label{Kap:Results}

Tables \ref{Tab:Versuchsuebersicht_Ti6Al4V} and \ref{Tab:Versuchsuebersicht_Ck45} in the appendix list the process parameters for each experiment ID and indicates for which experiments chip thicknesses are measured and for which etched samples are documented. The complete cutting test results are provided for \Titan in table \ref{Tab:TestResults_Ti6Al4Vkpl} and for Ck45 in table \ref{Tab:TestResults_Ck45kpl} in the \ref{Kap:Anhang_Meszdaten}. The process forces $F_c$ and $F_f$ are normalized to a cutting width of $w=1mm$ and are given together with their respective standard deviations $\sigma_{F_c}$ and $\sigma_{F_f}$. The next two columns contain the averaged cutting edge radius $r_n$ and its standard deviation $\sigma_{r_n}$, followed by the average chip thickness $h_{avg}$ and its standard deviation $\sigma_{h_{avg}}$ (if more than one chip is evaluated), the cutting distance $l_{cut}$ and the temperature of the chip $T_{chip}$ derived from the tempering color (only Ck45). The last two columns contain a qualitative statement of the tool wear and the quality of the process forces measurement.

\subsection{Process Forces}

\begin{figure}[h]
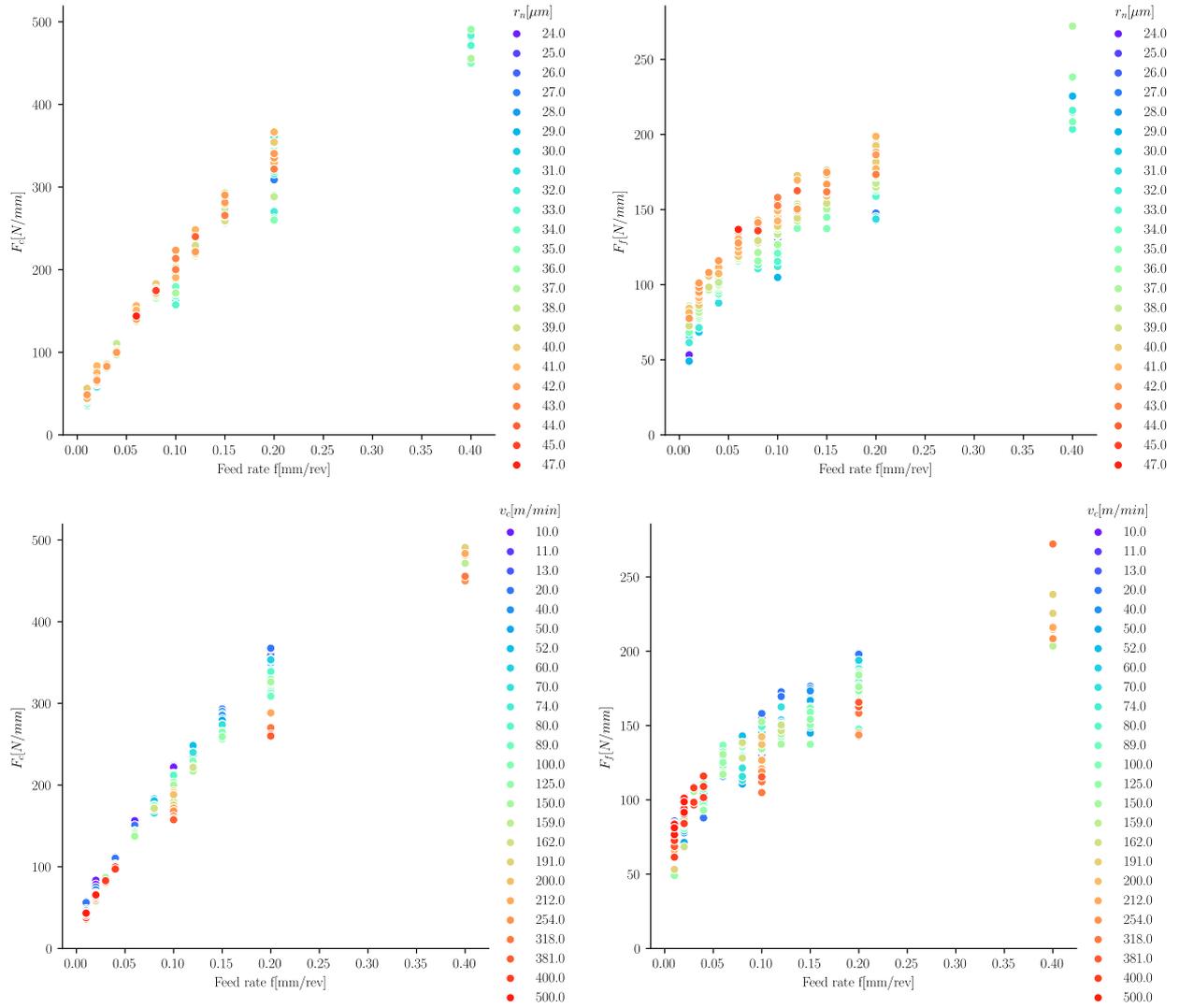

	\includesvg[width=0.49\textwidth]{Bilder/Ti6Al4V_pairplot_f_Fc_rn}
	\includesvg[width=0.49\textwidth]{Bilder/Ti6Al4V_pairplot_f_Ff_rn}
	\includesvg[width=0.49\textwidth]{Bilder/Ti6Al4V_pairplot_f_Fc_vc}
	\includesvg[width=0.49\textwidth]{Bilder/Ti6Al4V_pairplot_f_Ff_vc}	
	\caption{Process forces \Titan cutting experiments, cutting (left column) and feed force component (right column) depending on the feed $f$, the colour depicts the cutting edge radius $r_n$ (top row) and the cutting speed $v_c$ (bottom row) of the experiment.}
	\label{Bild:Ti6Al4V_Prozeszkraefte}
\end{figure}

\begin{figure}[h]
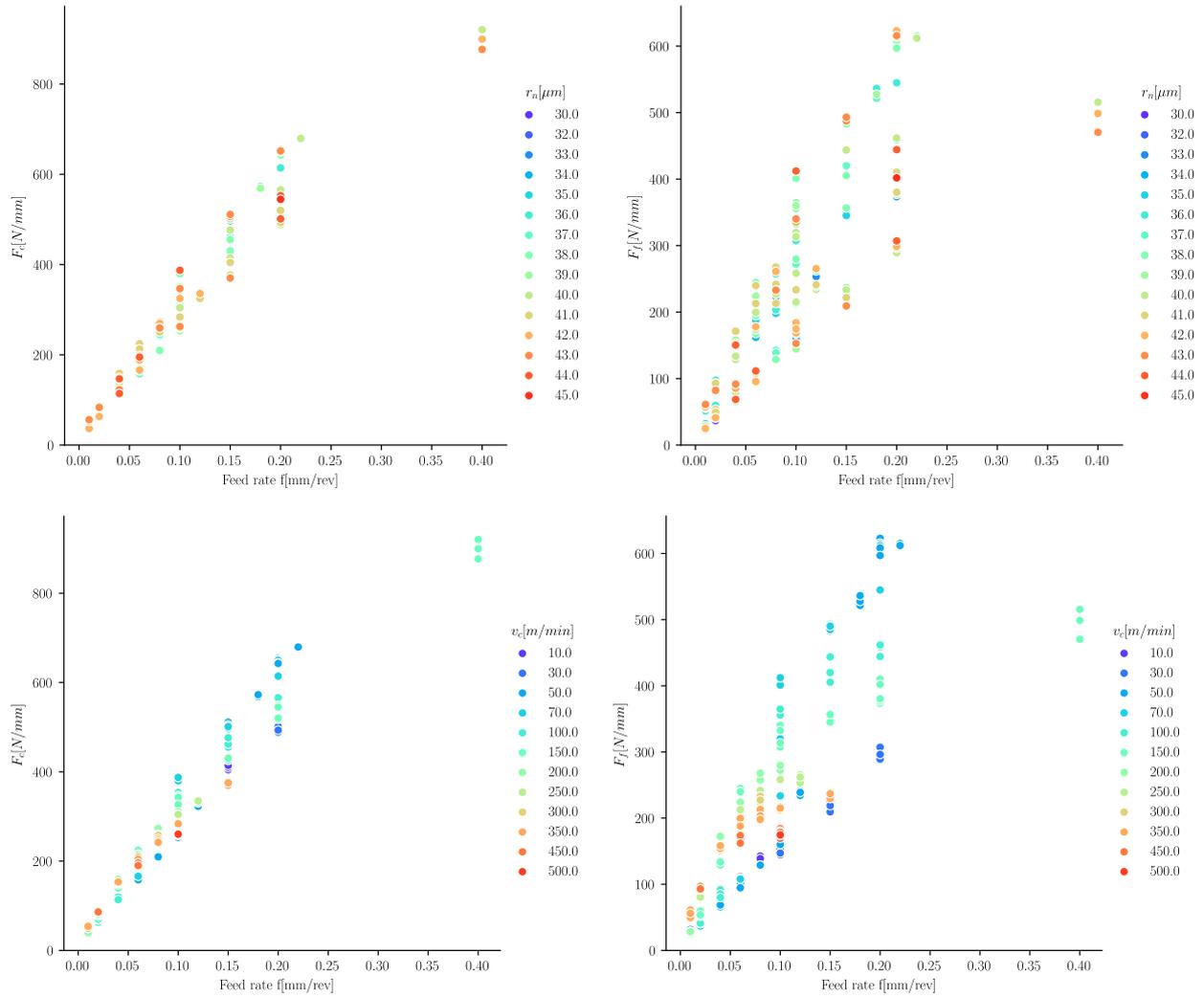

	\includesvg[width=0.49\textwidth]{Bilder/Ck45_pairplot_f_Fc_rn}
	\includesvg[width=0.49\textwidth]{Bilder/Ck45_pairplot_f_Ff_rn}
	\includesvg[width=0.49\textwidth]{Bilder/Ck45_pairplot_f_Fc_vc}
	\includesvg[width=0.49\textwidth]{Bilder/Ck45_pairplot_f_Ff_vc}
	\caption{Process forces Ck45 cutting experiments, cutting (left column) and feed force component (right column) depending on the feed $f$, the colour depicts the cutting edge radius $r_n$ (top row) and the cutting speed $v_c$ (bottom row) of the experiment.}
	\label{Bild:Ck45_Prozeszkraefte}
\end{figure}

The experimentally measured process forces are displayed in in Figure \ref{Bild:Ti6Al4V_Prozeszkraefte} for \Titan and in Figure \ref{Bild:Ck45_Prozeszkraefte} for Ck45. The colour of the dots indicates the cutting edge radius $r_n$ or cutting speed $v_c$ of the respective experiment. For \TitanPunkt, cutting and feed force increase with increasing feed. The cutting forces $F_c$ tend to increase with increasing cutting edge radius $r_n$ but decrease towards higher cutting speed $v_c$. The feed force $F_f$ increases as well with increasing $r_n$, but increasing $v_c$ leads to higher forces only until feeds around $f=\SI{0,1}{\milli\meter}$.

The Ck45 exhibits a complex behaviour: the process forces increase as well with increasing feed but dependencies on the cutting edge radius $r_n$ or cutting speed $v_c$ are not obvious. Instead, at a constant feed the process forces first increase from very low to medium cutting speeds ($v_c \approx \SI{100}{\meter/\minute}$) and then start to drop to similar low forces as with very low cutting speeds, see for example the feed forces at $f=\SI{0,1}{\milli\meter}$ in Figure \ref{Bild:Ck45_Prozeszkraefte}.

\FloatBarrier

\subsection{Chip Shapes}
\label{Kap:Spantypen_Spanformen}


At least one chip sample of every feed rate and cutting speed combination is etched. In the following a small selection is shown for \Titan and Ck45. For \TitanPunkt, chips in a feed range of $f=0.01mm, 0.1mm$ and $0.2mm$ and at cutting speeds of  $v_c=11, 40, 191$ and $381m/min$ are displayed in table \ref{Tab:Ti6Al4V_Spanueberblick}. All chips show chip segmentation where towards higher feed $f$ and higher cutting speed $v_c$ the segmentation becomes more pronounced. In all \Titan cutting experiments chip segmentation occurred, but traces of the effect of built-up edges (BUE) cannot be identified on any of the chips.

\begin{table}[h]
	\footnotesize
	\centering
		\begin{tabular}{cccc}
			$f [mm/rev]$ & 0.01 & 0.1 & 0.2\\
			$v_c [m/min]$\\
			11 & \raisebox{-0.5\totalheight}{\includegraphics[width=45mm]{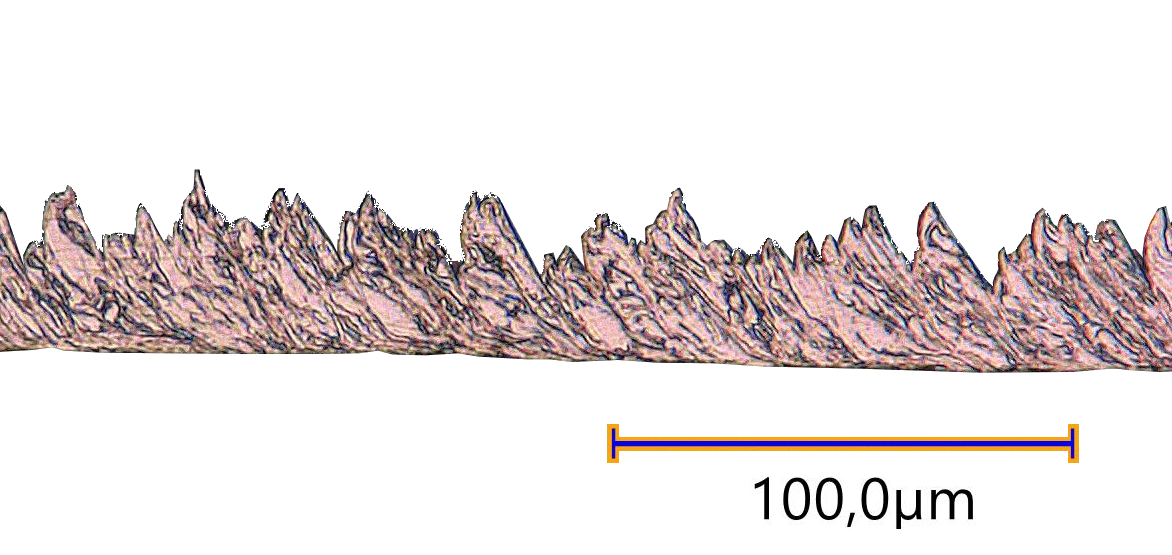}}
			& \raisebox{-0.5\totalheight}{\includegraphics[width=45mm]{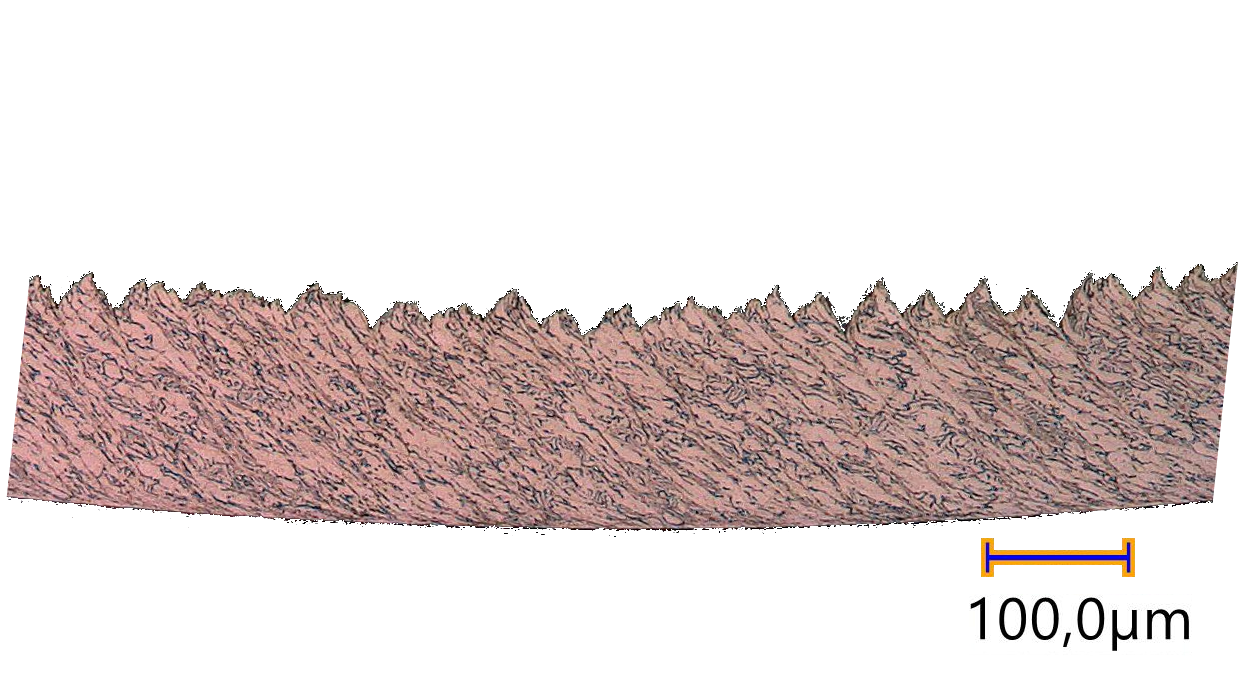}}
			& \raisebox{-0.5\totalheight}{\includegraphics[width=45mm]{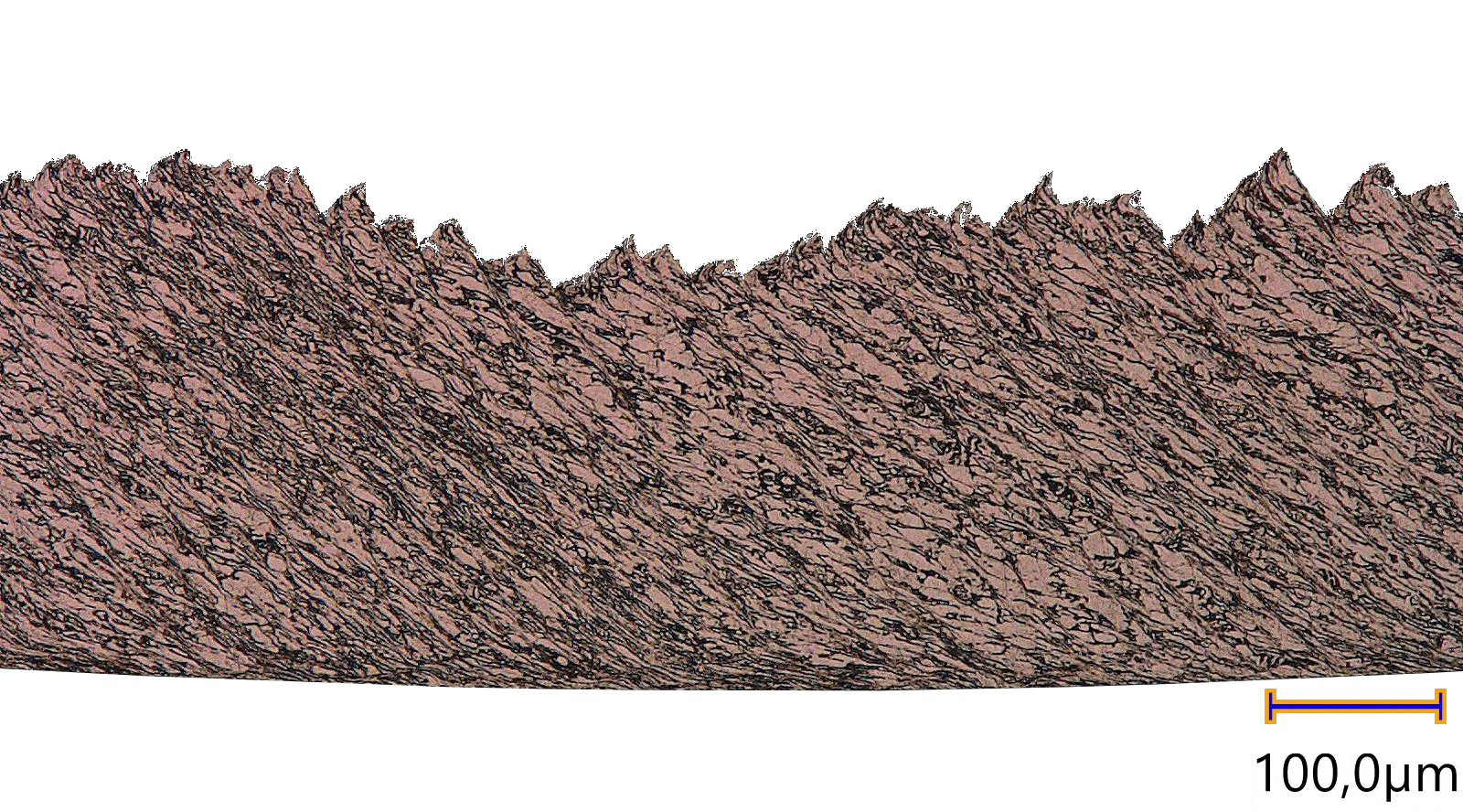}}\\
			\\
			40 & \raisebox{-0.5\totalheight}{\includegraphics[width=45mm]{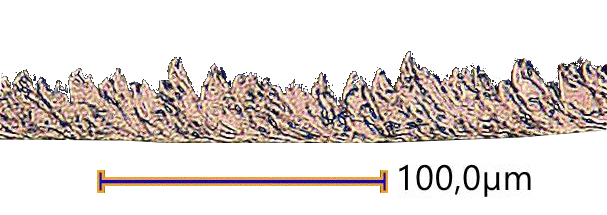}} 
			& \raisebox{-0.5\totalheight}{\includegraphics[width=45mm]{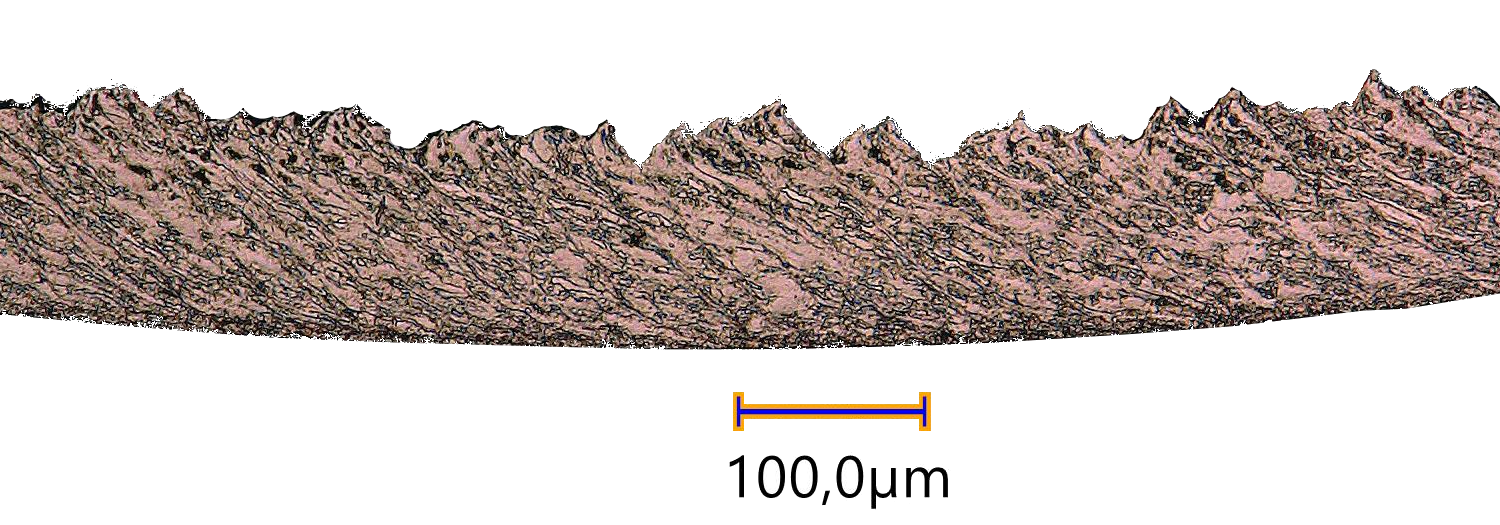}}
			& \raisebox{-0.5\totalheight}{\includegraphics[width=45mm]{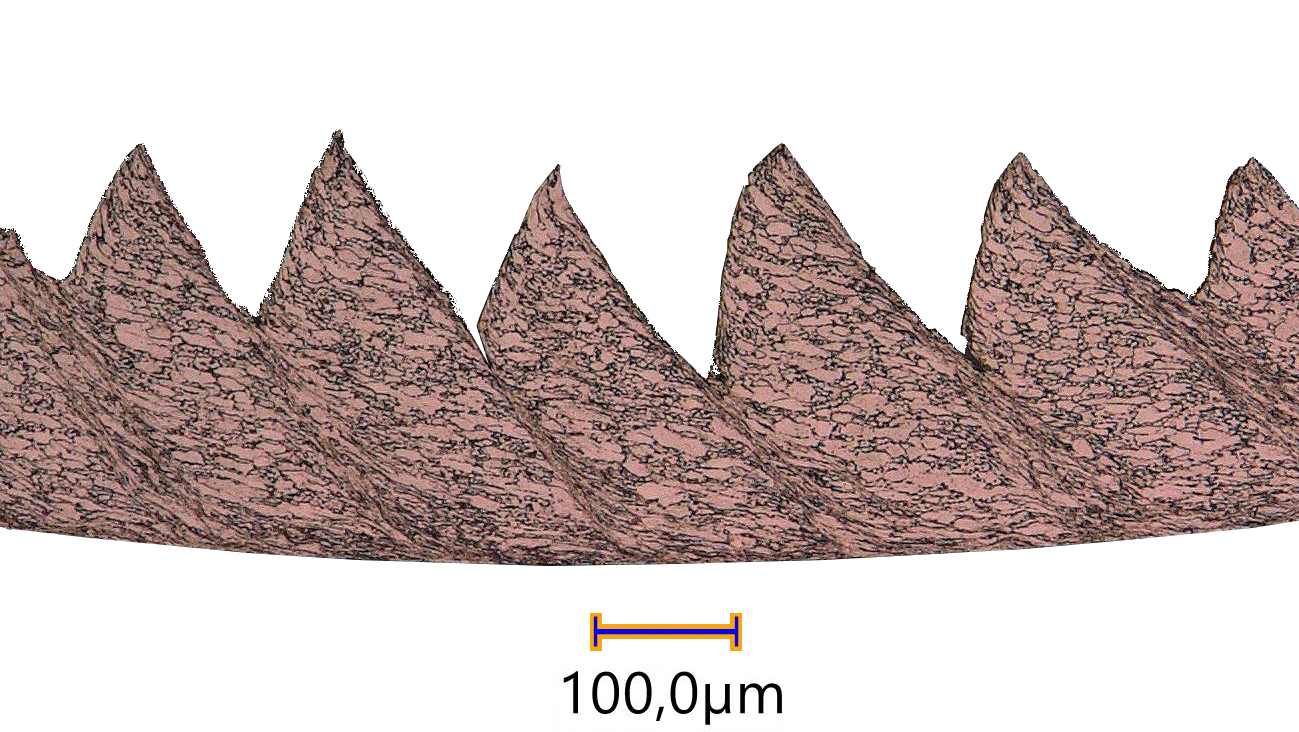}}\\
			\\
			100 & \raisebox{-0.5\totalheight}{\includegraphics[width=45mm]{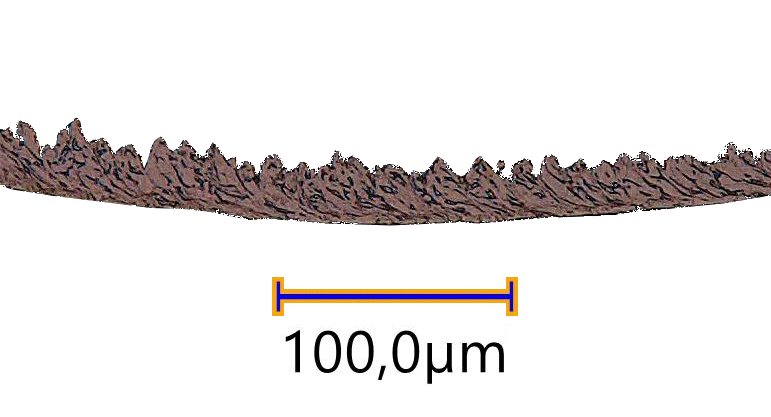}}
			& \raisebox{-0.5\totalheight}{\includegraphics[width=45mm]{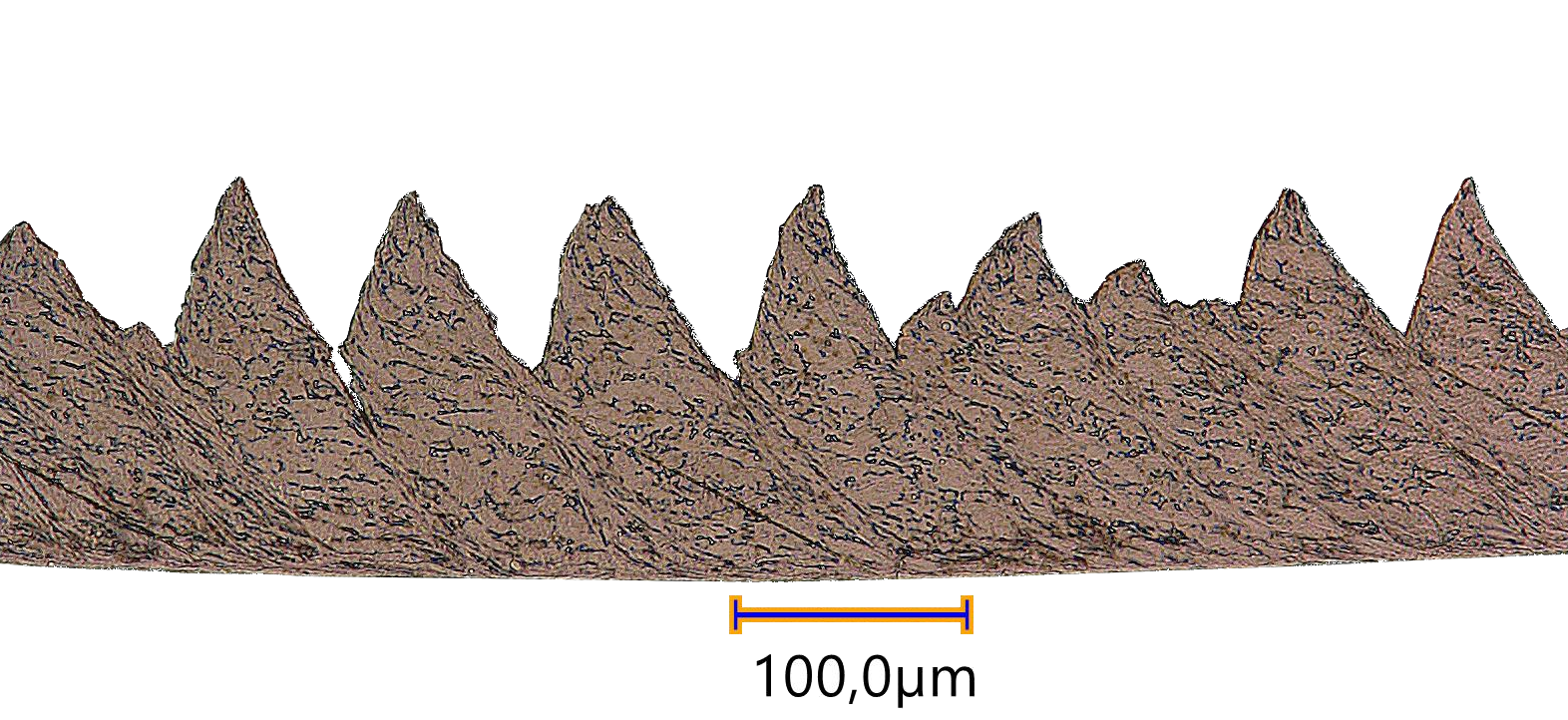}}
			& \raisebox{-0.5\totalheight}{\includegraphics[width=45mm]{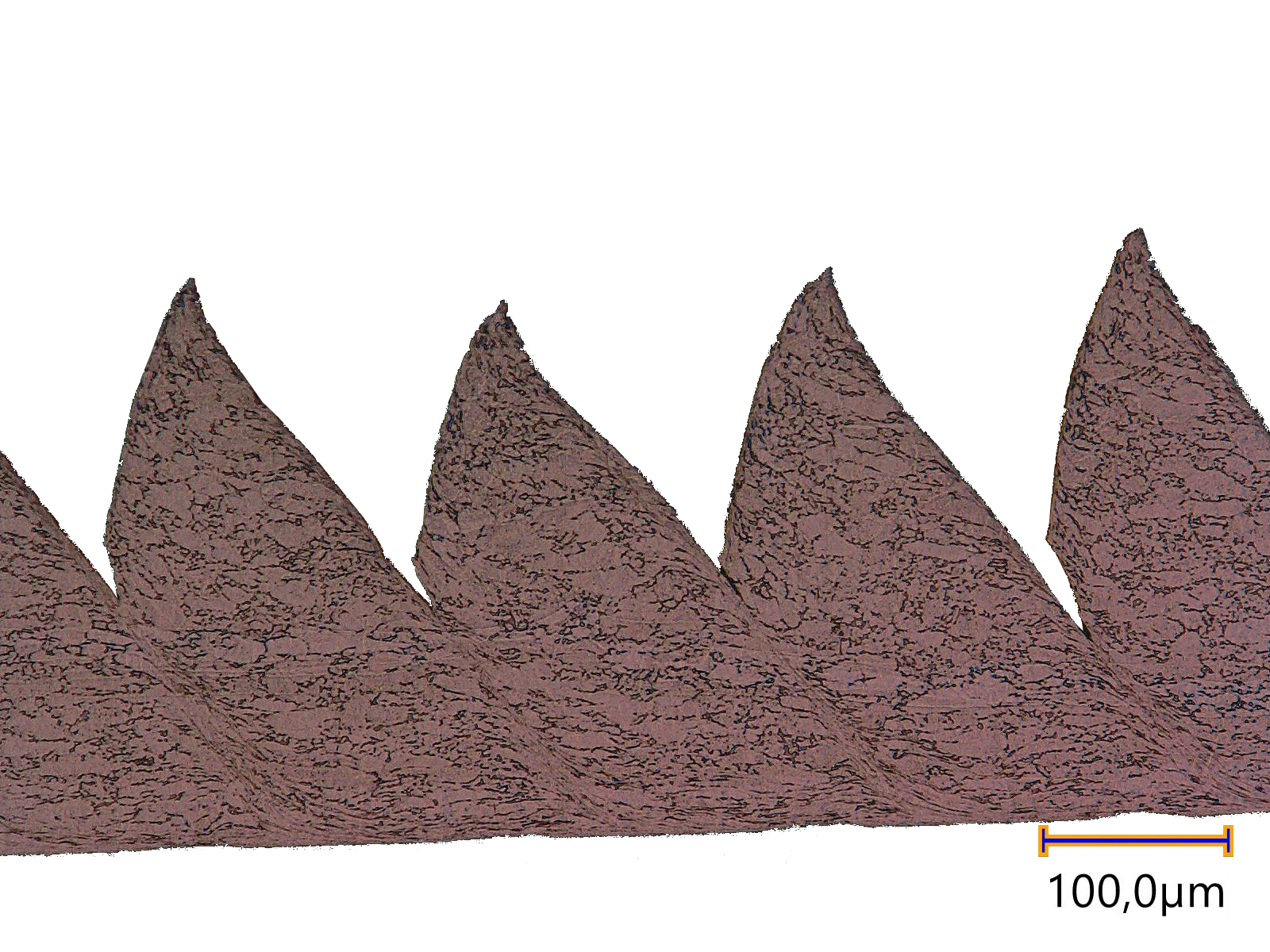}}\\
			\\
			381 & \raisebox{-0.5\totalheight}{\includegraphics[width=45mm]{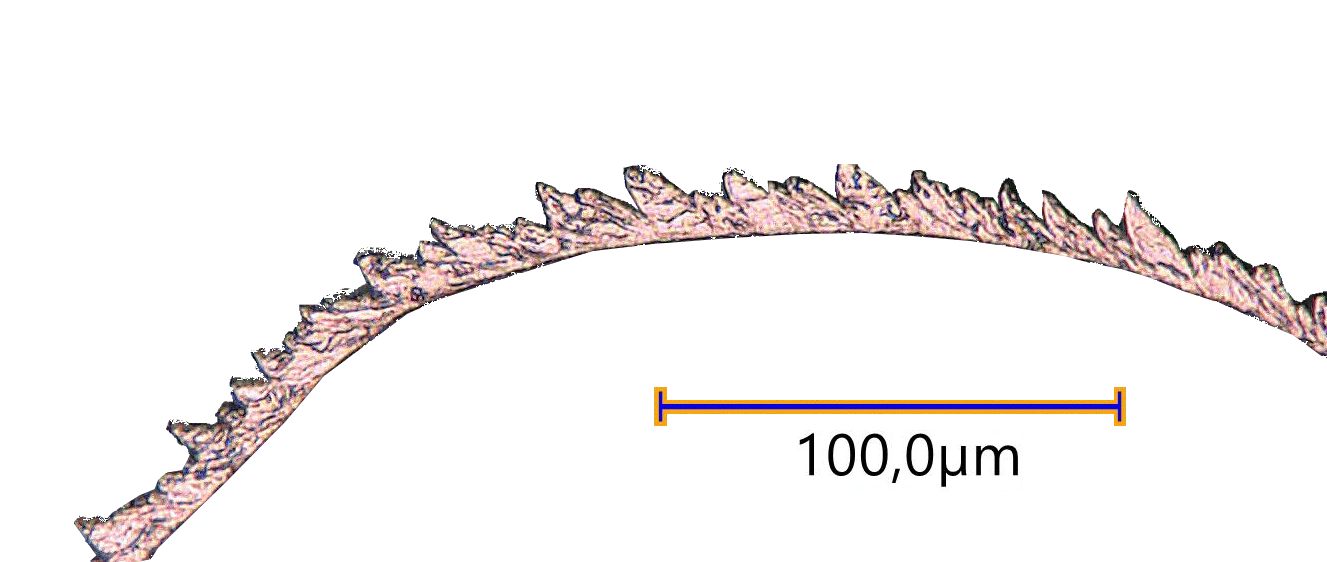}}
			& \raisebox{-0.5\totalheight}{\includegraphics[width=45mm]{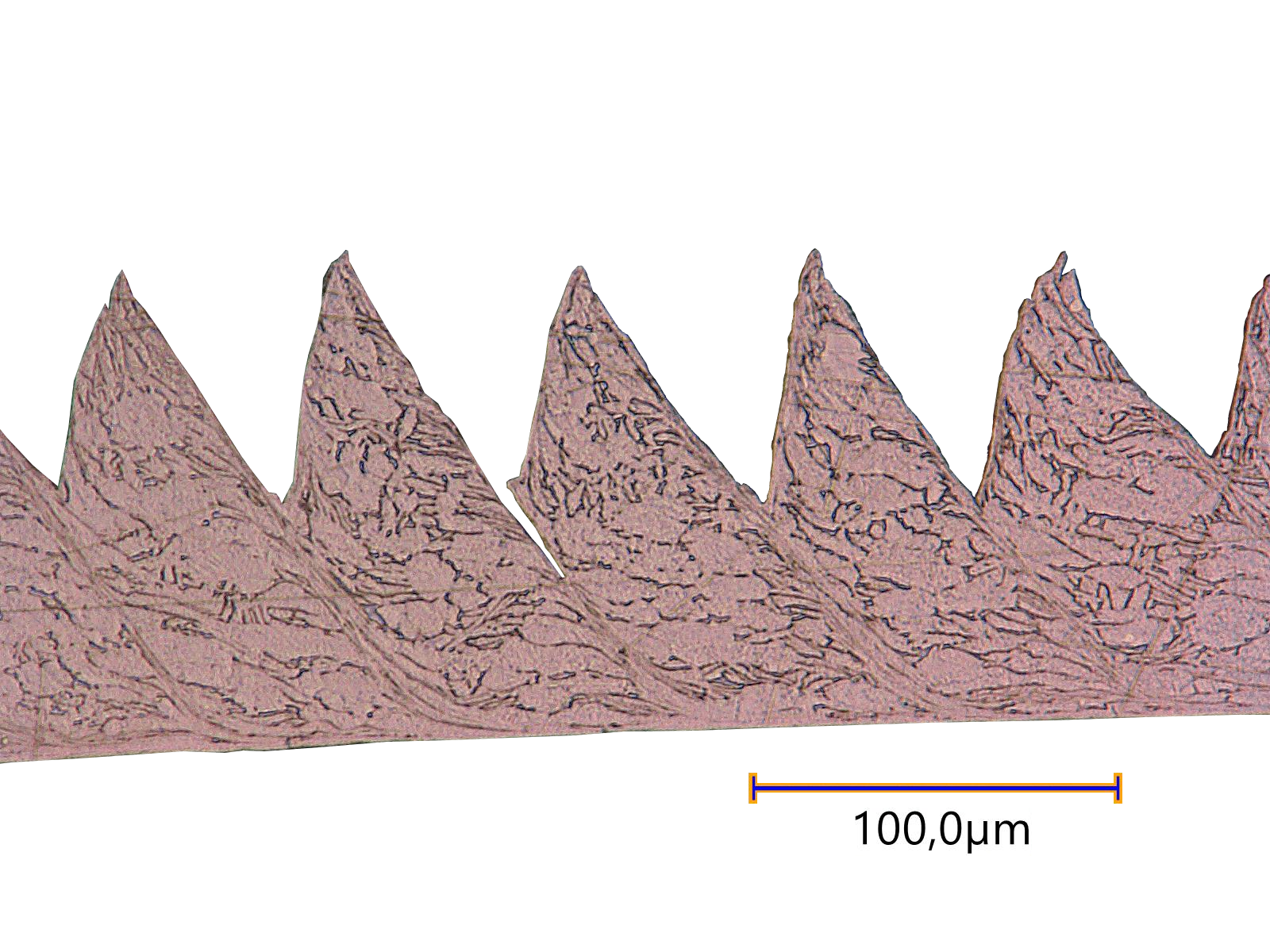}}
			& \raisebox{-0.5\totalheight}{\includegraphics[width=45mm]{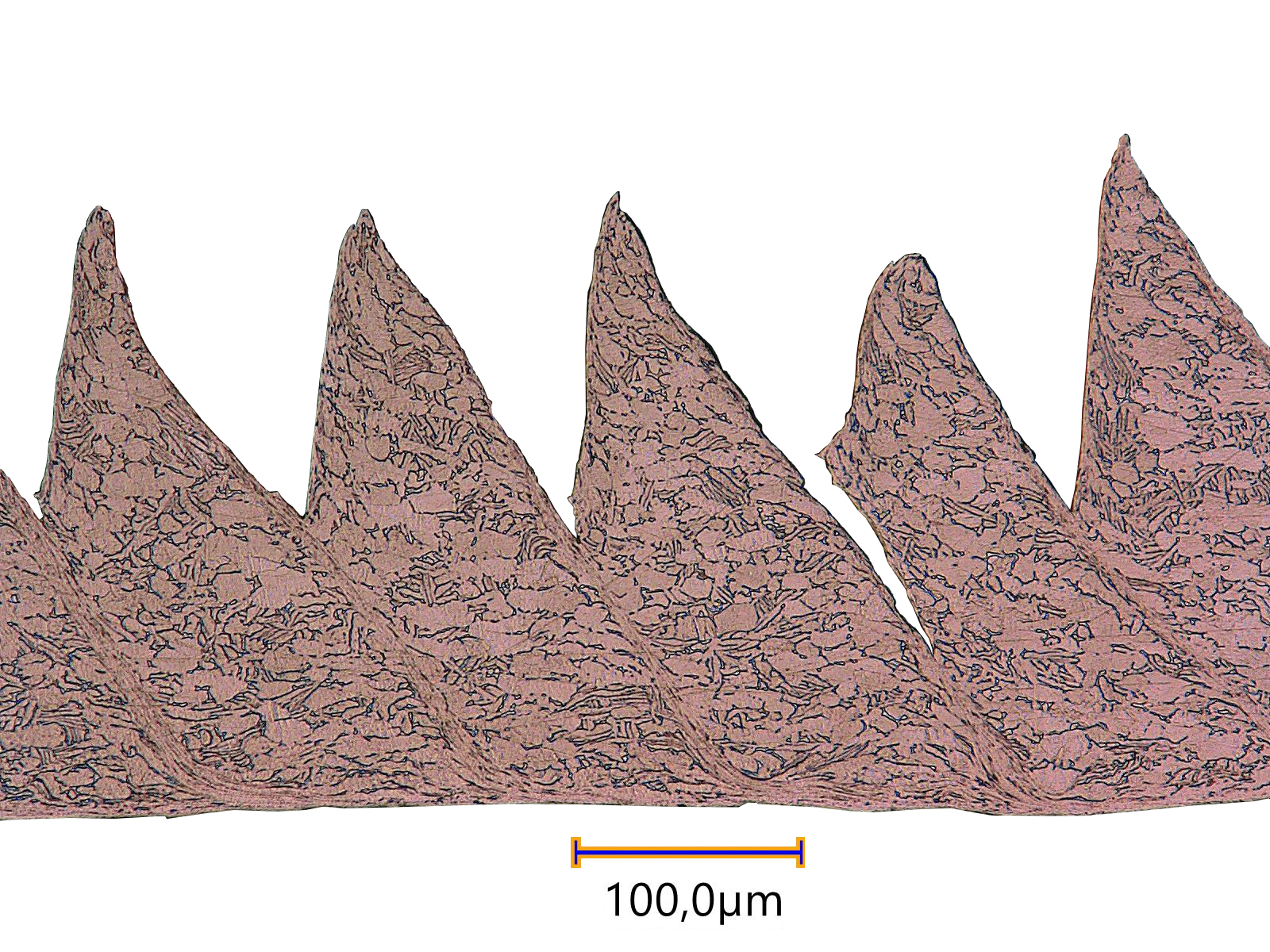}}\\
		\end{tabular}
	\normalsize
	\caption{\TitanPunkt: Chip overview of selected experiments with the chip flow direction in all images from right to left. Towards higher cuttings speed and higher feed rates the chip segmentation becomes more pronounced.}
	\label{Tab:Ti6Al4V_Spanueberblick}
\end{table}

\begin{figure}[h]
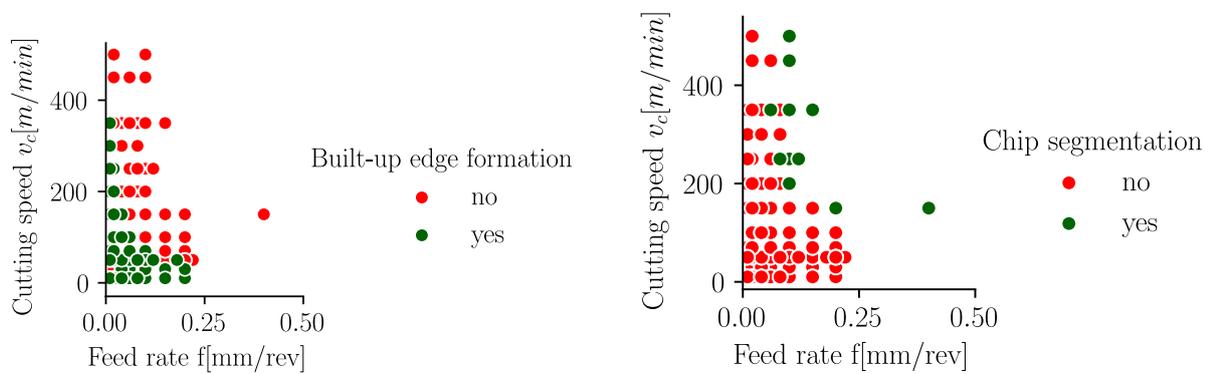

	\includesvg[width=0.49\textwidth]{Bilder/Ck45_BUE}
	\includesvg[width=0.49\textwidth]{Bilder/Ck45_Spansegmentierung}
	\caption{Built-up edge formation (left) and chip segmentation (right) in Ck45 depending on cutting speed $v_c$ and feed $f$.}
	\label{Bild:Ck45_BUE_Segmentierung}
\end{figure}

For Ck45, chips in a feed range of $f=0.02mm, 0.06mm$ and $0.1mm$ and at cutting speeds of  $v_c=10, 50, 150$ and $450m/min$ are displayed in table \ref{Tab:Ck45_Spanueberblick}. Ck45 chips show BUE traces in the chips up to $v_c=50m/min$ at all three feed levels, while at $v_c=150m/min$ only for $f=0.02mm$ BUE is visible and for $v_c=450m/min$ no BUE is visible at all. At higher cutting speeds and higher feeds of the Ck45 cutting experiments the BUE-formation diminishes, while chips segmentations start to form, see Figure \ref{Bild:Ck45_BUE_Segmentierung}.

\begin{table}[h]
	\footnotesize
	\centering
		\begin{tabular}{cccc}
			$f [mm/rev]$ & 0.02 & 0.06 & 0.1\\
			$v_c [m/min]$\\
			10 & \raisebox{-0.5\totalheight}{\includegraphics[width=45mm]{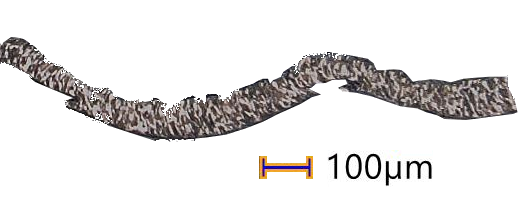}}
			& \raisebox{-0.5\totalheight}{\includegraphics[width=45mm]{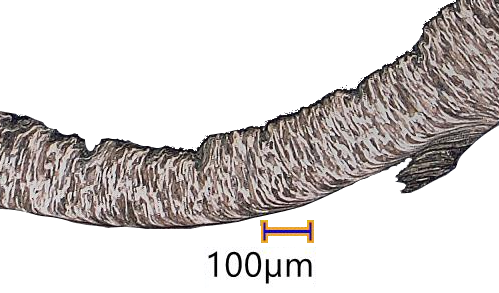}}
			& \raisebox{-0.5\totalheight}{\includegraphics[width=45mm]{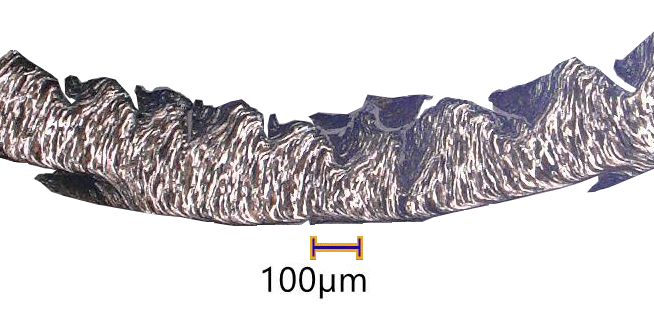}}\\
			\\
			50 & \raisebox{-0.5\totalheight}{\includegraphics[width=45mm]{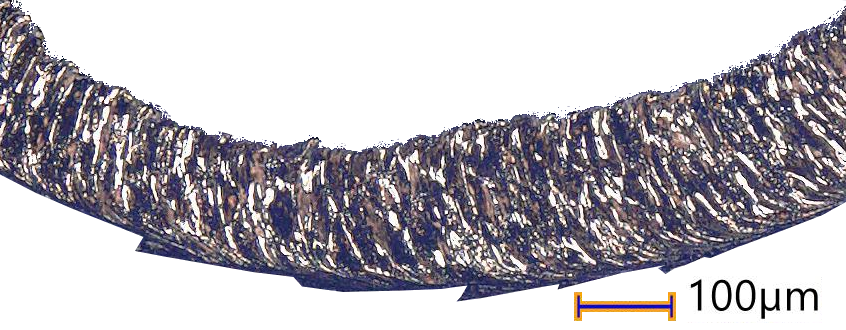}} 
			& \raisebox{-0.5\totalheight}{\includegraphics[width=45mm]{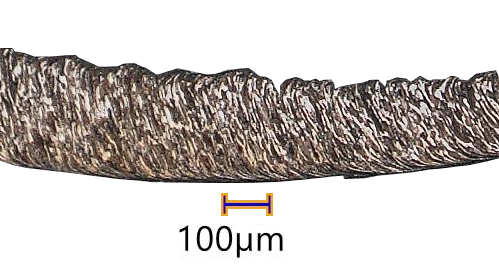}}
			& \raisebox{-0.5\totalheight}{\includegraphics[width=45mm]{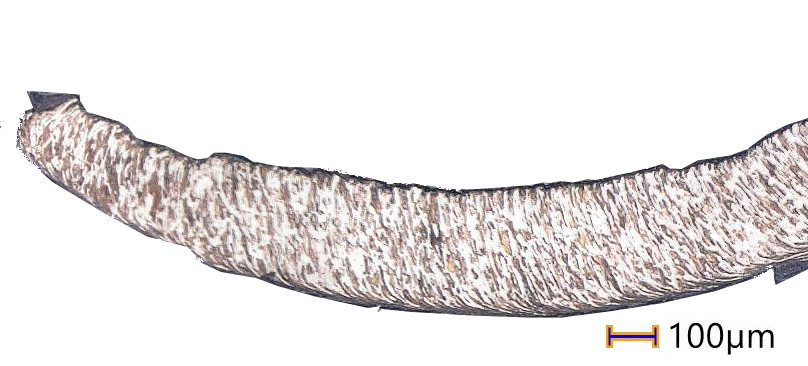}}\\
			\\
			150 & \raisebox{-0.5\totalheight}{\includegraphics[width=45mm]{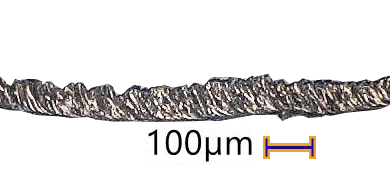}}
			& \raisebox{-0.5\totalheight}{\includegraphics[width=45mm]{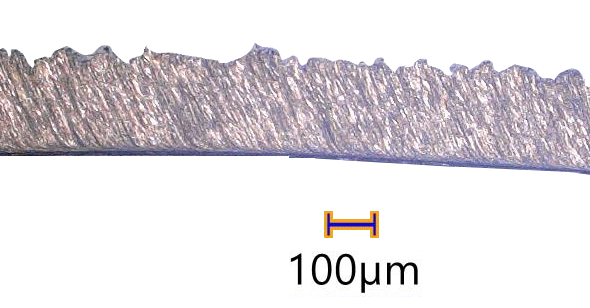}}
			& \raisebox{-0.5\totalheight}{\includegraphics[width=45mm]{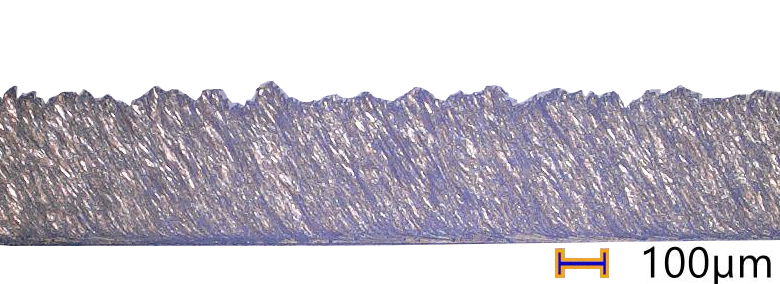}}\\
			\\
			450 & \raisebox{-0.5\totalheight}{\includegraphics[width=45mm]{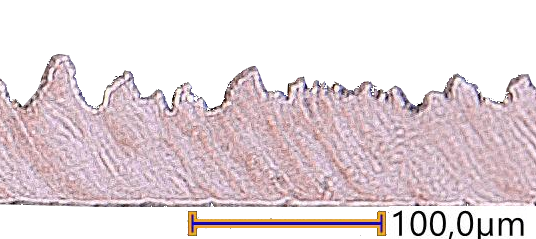}}
			& \raisebox{-0.5\totalheight}{\includegraphics[width=45mm]{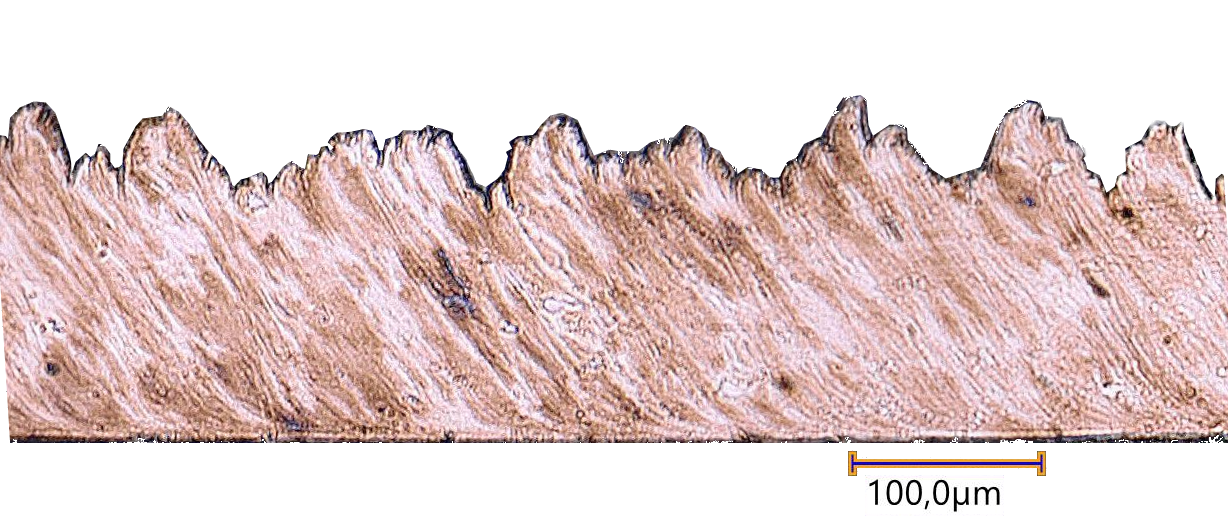}}
			& \raisebox{-0.5\totalheight}{\includegraphics[width=45mm]{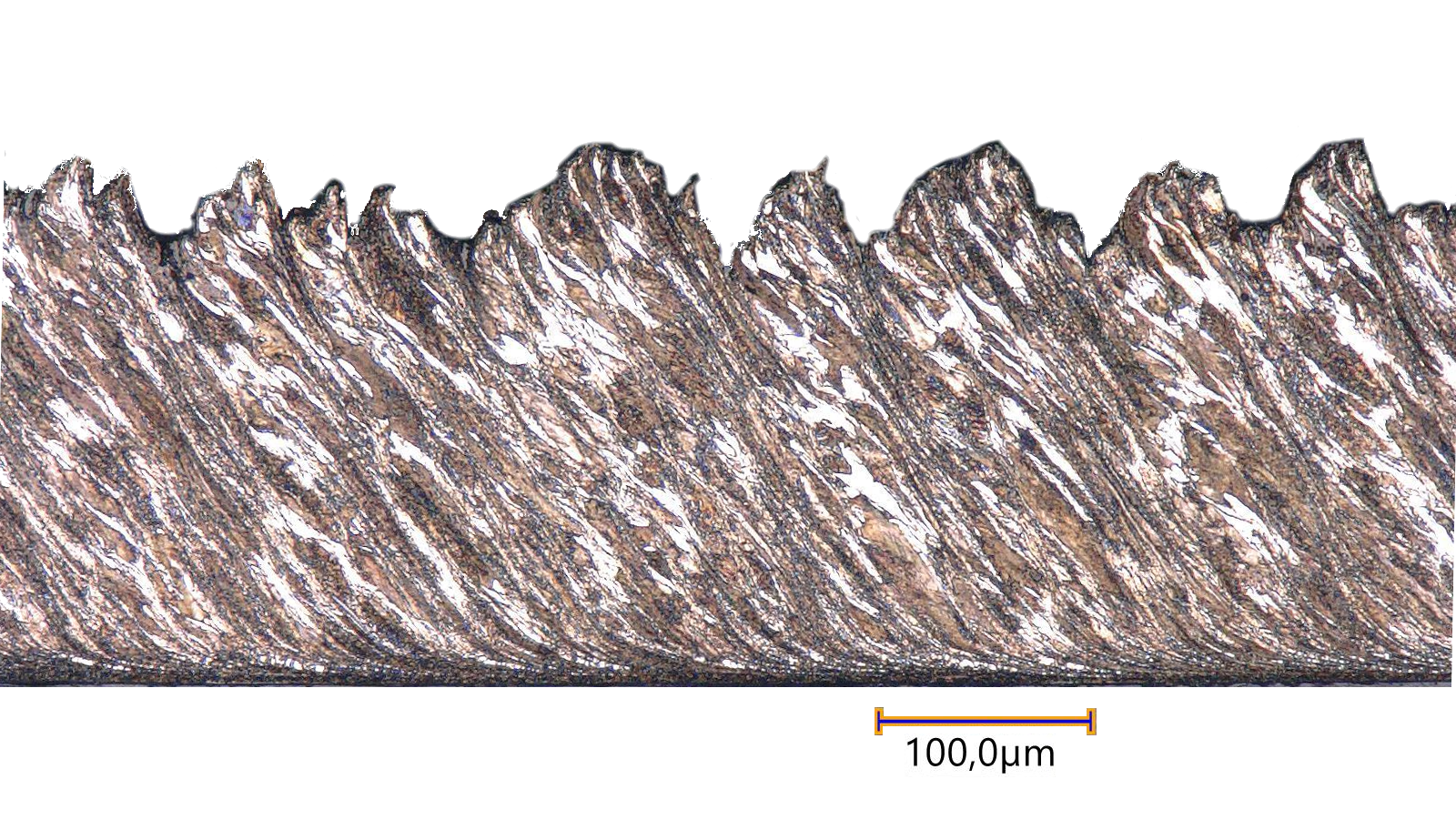}}\\
		\end{tabular}
	\normalsize
	\caption{Ck45: Chip overview of selected experiments with the chip flow direction in all images from right to left. Towards higher cuttings speed and higher feed rates chip segmentation starts to form, while traces of BUE diminish and disappear.}
	\label{Tab:Ck45_Spanueberblick}
\end{table}

\FloatBarrier


\subsection{Chip Thicknesses}
%

\begin{figure}[htp]
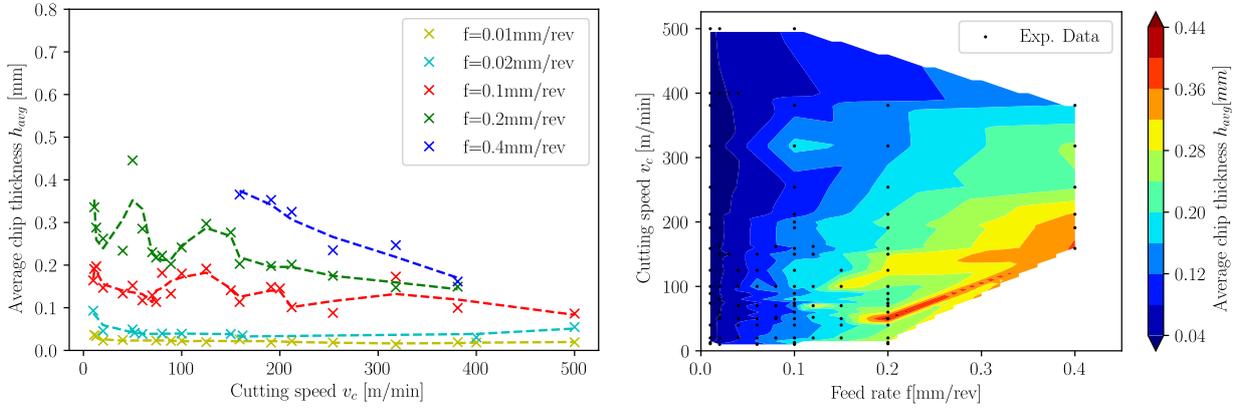

	\center{
		\includesvg[width=0.49\textwidth]{Bilder/Ti6Al4V_Spandicken_f_vc}
		\includesvg[width=0.49\textwidth]{Bilder/Ti6Al4V_Spandicken_f_vc_2D}
	}
	\caption{Measured average chip thicknesses of \Titan for different feeds $f$ and varying cuttings speeds $v_c$. Evaluation for selected feed rates (left) and interpolated contour plot (right).}
	\label{Bild:Spandicken_Ti6Al4V}
\end{figure}

\begin{figure}[htp]
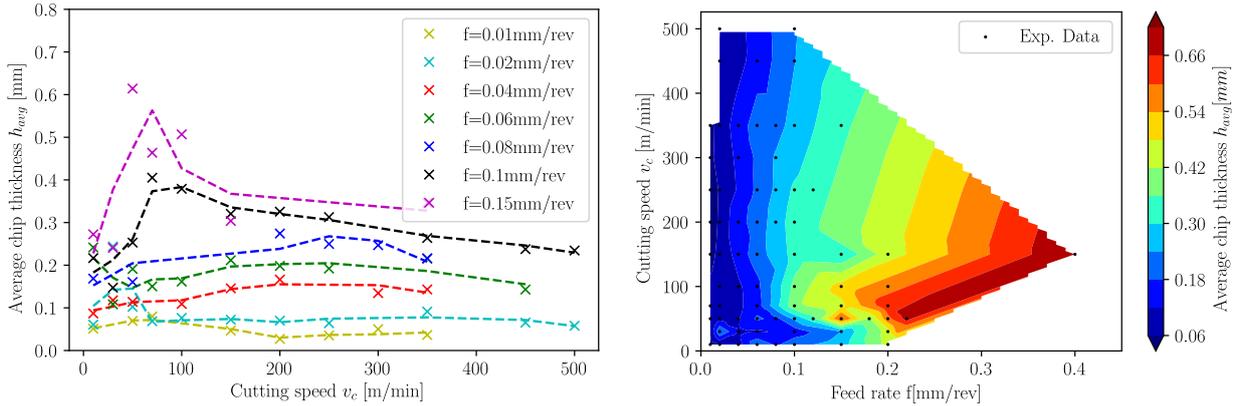

	\center{
		\includesvg[width=0.49\textwidth]{Bilder/Ck45_Spandicken_f_vc}
		\includesvg[width=0.49\textwidth]{Bilder/Ck45_Spandicken_f_vc_2D}
	}
	\caption{Measured average chip thicknesses of Ck45 for different feeds $f$ and varying cuttings speeds $v_c$. Evaluation for selected feed rates (left) and interpolated contour plot (right).}
	\label{Bild:Spandicken_Ck45}
\end{figure}

At least one chip of each combination of feed and cutting speed is embedded in Bakelite for the manual measurement of $h_{avg}$ using the method described in chapter \ref{Kap:Spandickenmessung}. The measured thicknesses are documented in the result tables \ref{Tab:TestResults_Ti6Al4Vkpl} (\TitanPunkt) and \ref{Tab:TestResults_Ck45kpl} (Ck45). The average chip thicknesses show high scatter for \Titan and Ck45. For this reason trendlines are created for some selected feeds using the Savitzky-Golay filter \cite{Savitzky1964} in SciPy \cite{SciPy2020}. A trend of decreasing chip thicknesses towards an increase of the cutting speed $v_c$ at constant feed rate $f$ is visible for both, \Titan and Ck45, see Figures \ref{Bild:Spandicken_Ti6Al4V} and \ref{Bild:Spandicken_Ck45}, respectively. In case of Ck45, an initial increase of the average chip thickness with increasing cutting speed for feed rates $f \geq \SI{0,04}{\milli\meter}$ is seen. However, upon further increase of the cutting speed, the average chip thickness decreases again. The cutting speed at which the behaviour reverses is not constant for all feed rates: as the feed increases, the maximum of the average chip thickness moves towards smaller cutting speeds $v_c$.

The measured chip thicknesses show some scatter, the reasons could be due to:

\begin{itemize}
	\item the manual measurement procedure,
	\item misalignments of chips in the embedding procedure because the chips are curled not only in feed direction but also due to the cylinder radius and
	\item low stiffness of thinner chips which could lead to skewing and therefore larger effective cross-sections after embedding.
\end{itemize}

The low stiffness of thinner chips prevented as well from the large scale evaluation of chip curling radii.

\subsection{Friction Coefficient}

\begin{figure}[htp]
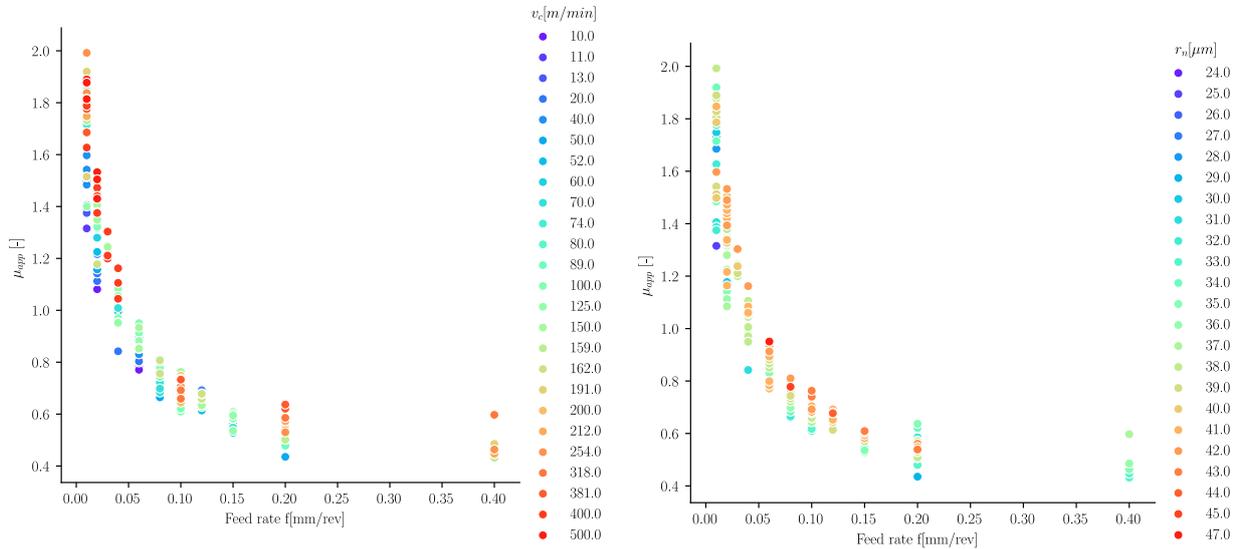

	\center{
		\includesvg[width=0.49\textwidth]{Bilder/Ti6Al4V_pairplot_f_mueapp_vc}
		\includesvg[width=0.49\textwidth]{Bilder/Ti6Al4V_pairplot_f_mueapp_rn}
	}
	\caption{Apparent friction coefficients $\mu_{app}$ versus feed $f$ for \Titan. The colour indicates the cutting speed $v_c$ (left) and the cutting edge radius $r_n$ (right). $\mu_{app}$ decreases with increasing feed $f$ and decreasing cutting speed $v_c$.}
	\label{Bild:Ti6Al4V_Reibung_Merchant}
\end{figure}

The measured process forces in the cutting experiments can be used to deduce friction coefficients. The simplest estimation is based on the ratio of tangential and normal forces exerted on the tool, also known as apparent friction coefficent $\mu_{app}$. It is given according to Merchant \cite{Merchant1945a,Merchant1945b} as:

\begin{equation}
	\label{Glg:Reibkoeffizient_Merchant}
	\mu_{app} = \frac{F_f + F_c \cdot tan \gamma}{F_c - F_f \cdot tan \gamma}
\end{equation}

with $F_f$ and $F_c$ being the feed and cutting components of the process forces and $\gamma$ being the rake angle. The drawback of the apparent friction coefficient $\mu_{app}$ is the neglection of the cutting edge radius influence assuming an ideal sharp tool (perfect wedge). The apparent friction coefficients are shown for \Titan cutting experiments in Figure \ref{Bild:Ti6Al4V_Reibung_Merchant}. Towards higher feeds the apparent friction coefficient decreases to levels of $\mu_{app} \approx 0.4..0.6$. Especially at very low feeds the cutting speed has a dominating influence on $\mu_{app}$ while with increasing $f$ the cutting edge radius $r_n$ becomes more important.

\begin{figure}[htp]
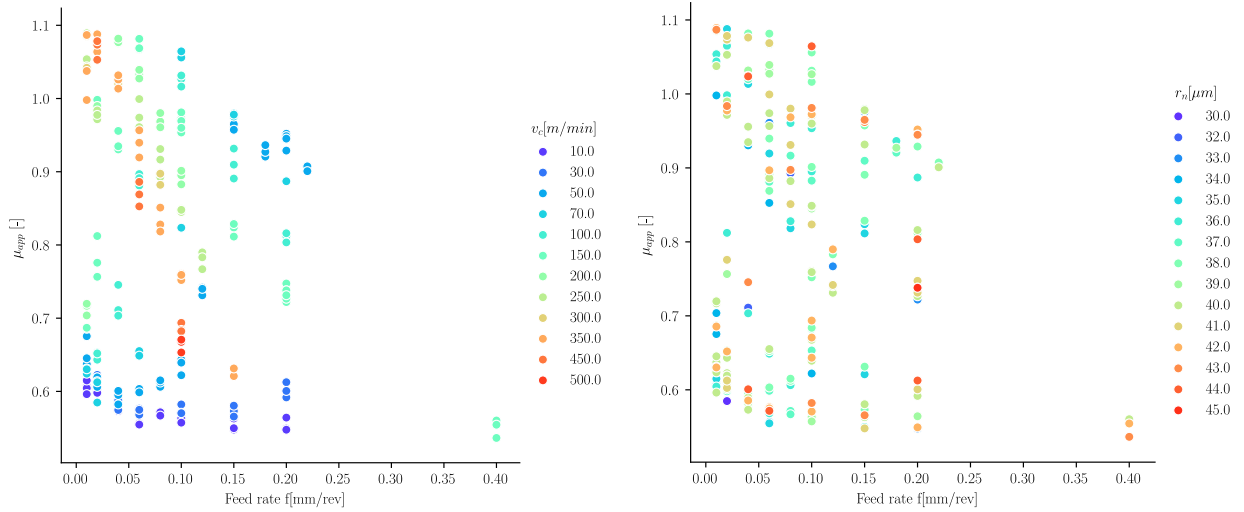

	\center{
		\includesvg[width=0.49\textwidth]{Bilder/Ck45_pairplot_f_mueapp_vc}
		\includesvg[width=0.49\textwidth]{Bilder/Ck45_pairplot_f_mueapp_rn}
	}
	\caption{Apparent friction coefficients $\mu_{app}$ versus feed $f$ for Ck45. The colour indicates the cutting speed $v_c$ (left) and the cutting edge radius $r_n$ (right). A dependency on the cutting edge radius $r_n$ is not visible, while for very low $f$ and increasing cutting speed $v_c$ an increase in $\mu_{app}$ is visible which then drops upon further increases in $v_c$. This behaviour is likely due to a change in BUE-formation and chip segmentation.}
	\label{Bild:Ck45_Reibung_Merchant}
\end{figure}

The apparent friction coefficients of the Ck45 cutting experiments are shown in Figure \ref{Bild:Ck45_Reibung_Merchant}. The cutting edge radius $r_n$ plays no obvious role in $\mu_{app}$, but for very low $f$ and increasing cutting speed $v_c$ an increase in $\mu_{app}$ is visible which then drops upon further increases in $v_c$. This behaviour is likely due to a change in BUE-formation and chip segmentation which changes towards increasing $f$ and $v_c$, see also Figure \ref{Bild:Ck45_BUE_Segmentierung}.

\FloatBarrier

\begin{figure}[htp]
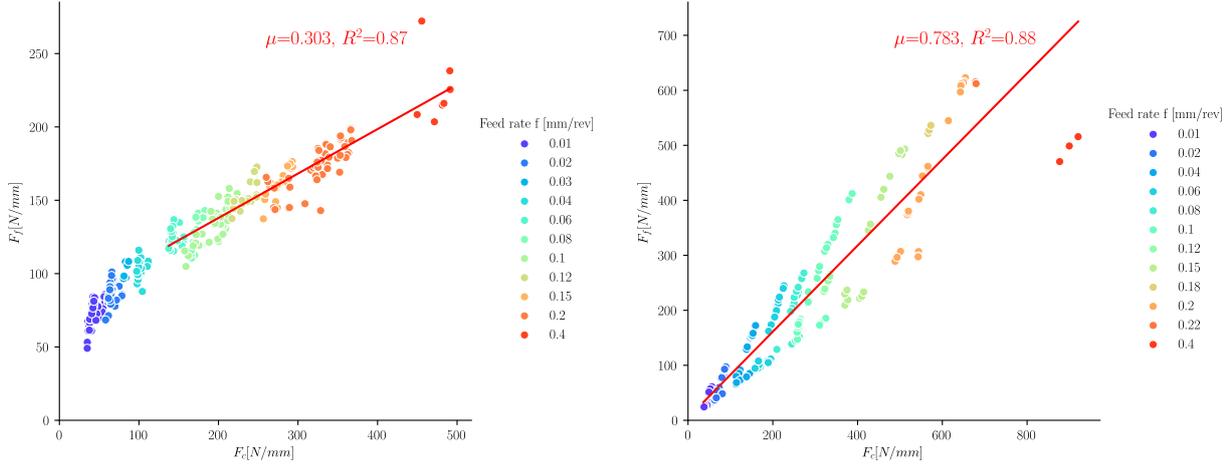

	\center{
		\includesvg[width=0.49\textwidth]{Bilder/Ti6Al4V_pairplot_Regression_Fc_Ff_f}
		\includesvg[width=0.49\textwidth]{Bilder/Ck45_pairplot_Regression_Fc_Ff_f}
	}
	\caption{Friction coefficient $\mu_{fr}$ estimation based on Albrecht \cite{Albrecht1960} for \Titan (left) and Ck45 (right). The estimated friction coefficient for Ck45 is very high and most likely inaccurate due to changes in BUE-formation and chip segmentation.}
	\label{Bild:Ti6Al4V_Reibung_Albrecht}
\end{figure}

A more elaborated approach for the determination of the friction coefficient from cutting experiments is described by Albrecht \cite{Albrecht1960}, considering the ploughing effect due to finite sharpness effect in the determination of the friction coefficient $\mu_{fr}$:

\begin{equation}
	\label{Glg:Reibkoeffizient_Albrecht}
	\mu_{fr} = \frac{(F_f-F_{pl,f}) + (F_c - F_{pl,c}) \cdot tan \gamma}{(F_c-F_{pl,c}) + (F_f - F_{pl,f}) \cdot tan \gamma}
\end{equation}

with $F_{pl,c}$ and $F_{pl,f}$ being the ploughing force components exerted in normal and feed direction. The friction coefficient $\mu_{fr}$ equals the slope in a $F_f-F_c$ - plot where at higher feeds ploughing effects diminish since the ratio of the cutting edge radius $r_n$ to the feed $f$ becomes very small. Using the measured process forces of all \TitanPunkt -experiments, the trend in Figure \ref{Bild:Ti6Al4V_Reibung_Albrecht} becomes visible from which the friction coefficient can be estimated as $\mu_{fr}=0.3$. Similarly, for Ck45 the friction coefficient is about $\mu_{fr}=0.78$, which is a very high value and most likely inaccurate due to changes in BUE-formation and chip segmentation. It is emphasized that a more credible deduction of friction coefficients requires cutting tests with specially prepared cutting edge radii $r_n$ as for example used in \cite{Wyen2011}. Alternatively, in-process tribometer can be used for the determination of the friction coefficient \cite{Meier2017} directly on the fresh cut surface.

\FloatBarrier

\subsection{Tempering Colours}

\begin{figure}[htp]
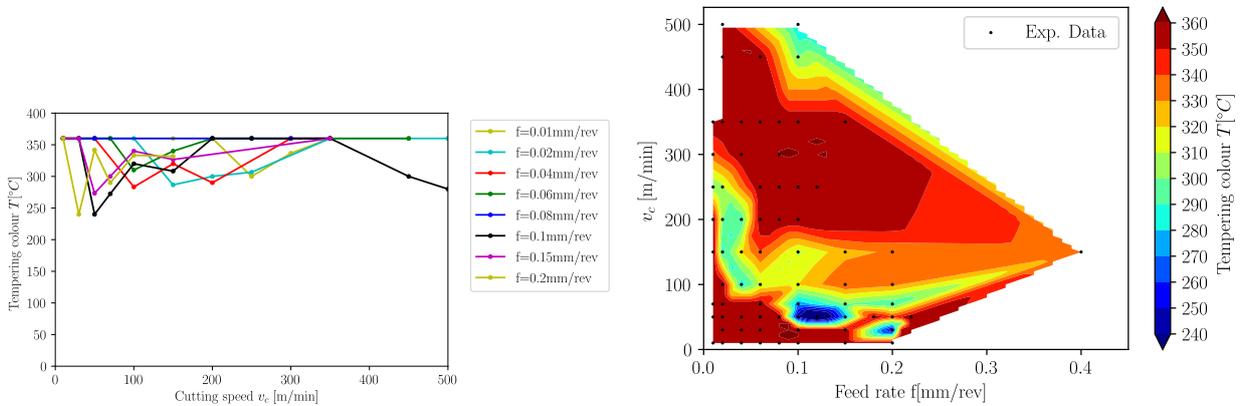

	\center{
		\includesvg[width=0.49\textwidth]{Bilder/Ck45_Anlaszfarben_f_vc}
		\includesvg[width=0.49\textwidth]{Bilder/Ck45_Anlaszfarben_f_vc_2D}
	}
	\caption{Estimated Ck45 chip temperatures based on tempering colours for different feeds $f$ and varying cuttings speeds $v_c$. Evaluation for selected feeds (left) and interpolated contour plot (right) with black dots denoting the experimental data points which have been used for the interpolation.}
	\label{Bild:Anlaszfarben_Ck45}
\end{figure}

During the cutting, the temperature in the chips increases due to plastic dissipation in the primary shear zone and friction in the secondary shear zone. In the case of the Ck45, these temperature increases lead to visible tempering colours in the chips, which can be matched to tabulated values \cite{Foerster2018}. The tempering colours can indicate temperatures as low as $T_{min}=200^\circ$ and as high as $T_{max}=360^\circ C$. All Ck45 chips are manually analysed and the tempering colour dependence on the cutting speed is shown for selected feeds together with interpolated contour plots in Figure \ref{Bild:Anlaszfarben_Ck45}. At low cutting speeds, temperatures are in the order of the maximum temperatur $T_{max}=360^\circ C$. With increasing cutting speed the temperatures first reduces to levels below $T=300^\circ C$ between $v_c=100..200m/min$. In cutting speed ranges of $v_c \approx 200..450m/min$ the temperatures increase to $T_{max}\approx 360^\circ$ followed by a temperature reduction when the cutting speed is increased beyond $v_c \approx 450m/min$. A possible explanation of this effect is a change in the chip formation mechanism see chapter \ref{Kap:Spantypen_Spanformen} and Figure \ref{Bild:Ck45_BUE_Segmentierung}.

\FloatBarrier

\subsection{Tool Wear}

The tool wear depends on the feed, cutting speed and the amount of material removed at these parameters. For each cutting experiment the tool wear is visually inspected and qualitatively classified into three categories:

\begin{itemize}
	\item Low (L): no or very few remains of material sticking to the rake and/or clearance face of the cutter
	\item Medium (M): beginning crater wear, material built-up on rake and/or clearance face
	\item High (H): heavy crater wear, cutting edge worn, large material built-up
\end{itemize}

\begin{figure}[htp]
	\center{
		\includegraphics[width=0.8\textwidth]{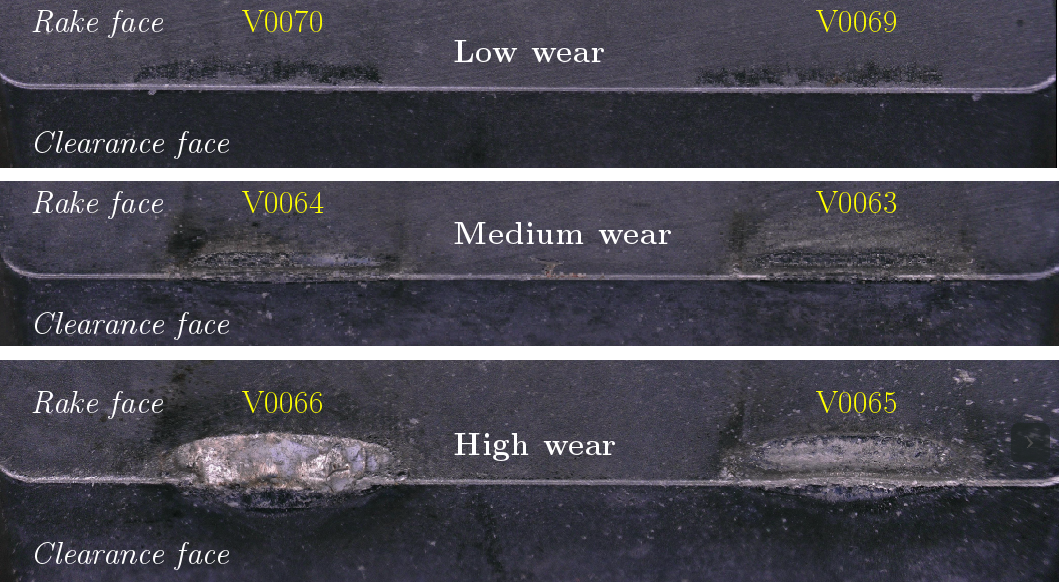}
	}
	\caption{Wear classification of the cutter inserts: examples for low (top), medium (middle) and high (bottom) wear. The text in yellow colour depicts the corresponding experiment number.}
	\label{Bild:WSP_Verschleiszstatus}
\end{figure}

Examples of different tool wear classifications are shown in \ref{Bild:WSP_Verschleiszstatus}. The tool wear classification is given for each experiment in the result tables \ref{Tab:TestResults_Ti6Al4Vkpl} (\TitanPunkt) and \ref{Tab:TestResults_Ck45kpl} (Ck45).

\FloatBarrier

\subsection{Kienzle Coefficients}


                                                                                                                                                                                                                                                                                                                          
The Kienzle-Victor process force model \cite{Kienzle1952} is an empirical equation that relates the cutting force $F_c$ to the cut width $b$, uncut chip thickness $h$ and a factor $k_c$ which is the cutting force per unit area: 

\begin{equation}
	F_c = k_c \cdot a_p \cdot f = k_c \cdot b \cdot h
\end{equation}

All effects of the process conditions like cutting speed, tool geometry as well as workpiece and tool material are contained in $k_c$. It was found by Kienzle that the cut width $b$ scales linearly with the cutting force while the uncut chip thickness scales with a power law, where $k_{c1.1}$ is the specific cutting force at $b=h=\SI{1}{\milli\meter}$:

\begin{equation}
	\label{Glg:KienzleVictor_Fc}
	F_c = k_{c1.1} \cdot h^{1-m_c}
\end{equation}

The model can be extended to to feed and passive force as well \cite{Koenig2008}:

\begin{eqnarray}
	\label{Glg:KienzleVictor_Ff}
	F_f & = & k_{f1.1} \cdot h^{1-m_f}\\
	\label{Glg:KienzleVictor_Fp}
	F_p & = & k_{p1.1} \cdot h^{1-m_p}
\end{eqnarray}

\begin{figure}[htp]
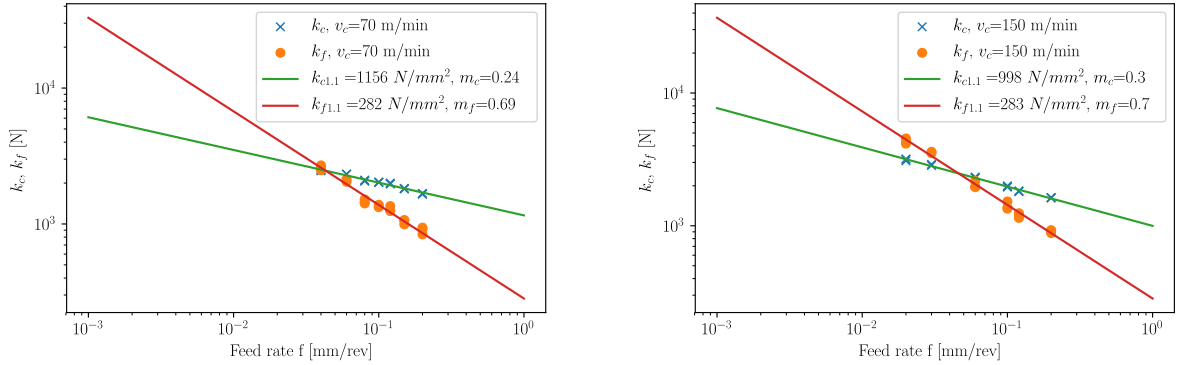

	\center{
		\includesvg[width=0.49\textwidth]{Bilder/Kienzle_Ti6Al4V_70m_min}
		\includesvg[width=0.49\textwidth]{Bilder/Kienzle_Ti6Al4V_150m_min}
	}
	\caption{Kienzle coefficient determination for \Titan at cutting speeds of $v_c=70m/min$ (left) and $v_c=150m/min$ (right).}
	\label{Bild:Kienzle_Ti6Al4V_vc70_150m_min}
\end{figure}

In quasi-orthogonal cutting the passive force component is $F_p \approx 0$ and therefore equations \eqref{Glg:KienzleVictor_Fc} and \eqref{Glg:KienzleVictor_Ff} are used to determine the Kienzle-coefficients $k_{c1.1}(v_c), k_{f1.1}(v_c)$ and the respective exponents $m_c(v_c), m_f(v_c)$ at different cutting speeds for \Titan and Ck45. For this purpose $k_c$ and $k_f$ are computed from the process forces and are extrapolated to $f=1mm$ for each cutting speed. Examples of such extrapolations are shown in Figure \ref{Bild:Kienzle_Ti6Al4V_vc70_150m_min} for two different cutting speeds.

\begin{figure}[htp]
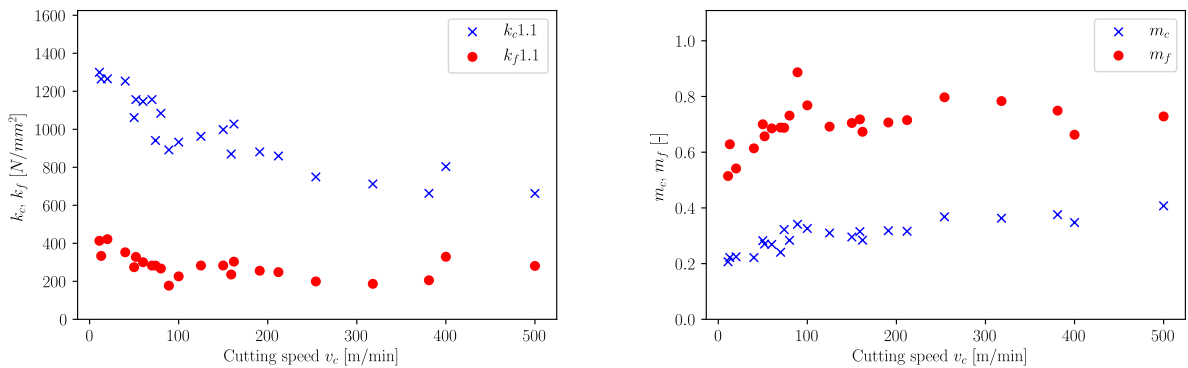

	\center{
		\includesvg[width=0.49\textwidth]{Bilder/Ti6Al4V_kc11_kf11_vc}
		\includesvg[width=0.49\textwidth]{Bilder/Ti6Al4V_zc11_zf11_vc}
	}
	\caption{Kienzle coefficients for \Titan and various cutting speeds $v_c$.}
	\label{Bild:Kienzle_Ti6Al4V}
\end{figure}

Figure \ref{Bild:Kienzle_Ti6Al4V} shows the dependencies of the Kienzle-coefficients on the cutting speed for \TitanPunkt. The corresponding values are supplied in table \ref{Tab:Kienzle_Ti6Al4V} in the \ref{Kap:Anhang_KienzleParameter}. Towards higher cutting speeds the specific cutting and feed force coefficients reduce, while the exponents $m_c, m_f$ increase.

\begin{figure}[htp]
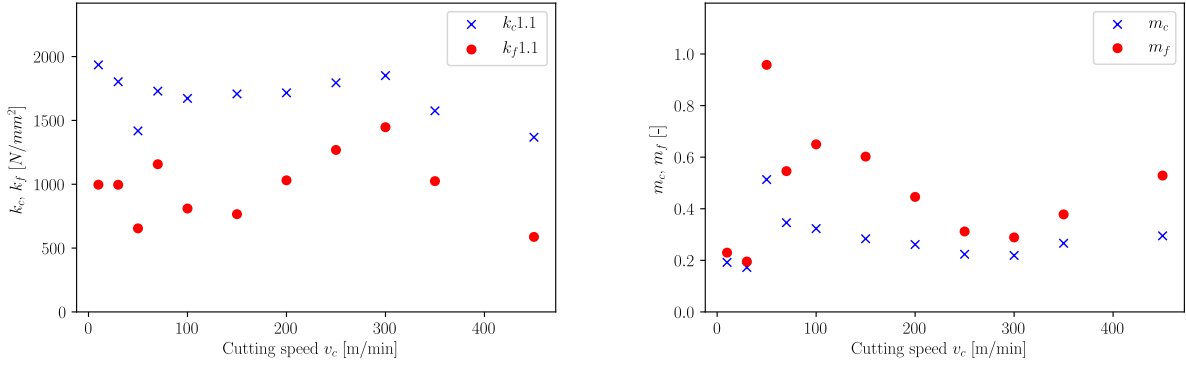

	\center{
		\includesvg[width=0.49\textwidth]{Bilder/Ck45_kc11_kf11_vc}
		\includesvg[width=0.49\textwidth]{Bilder/Ck45_zc11_zf11_vc}
	}
	\caption{Kienzle coefficients for Ck45 and various cutting speeds $v_c$.}
	\label{Bild:Kienzle_Ck45}
\end{figure}

For Ck45, the specific cutting and feed force coefficients slightly reduce until cutting speeds of $v_c = 150m/min$, followed by an increase towards $v_c = 300m/min$ and reduce from thereon strongly. The increase around $v_c = 300m/min$ is possibly induced by the dynamic strain aging (DSA) phenomenon \cite{Childs2019,Devotta2015,Devotta2020}. Figure \ref{Bild:Kienzle_Ck45} shows the dependencies of the Kienzle-coefficients on the cutting speed for Ck45. The corresponding values are supplied in table \ref{Tab:Kienzle_Ck45} in the \ref{Kap:Anhang_KienzleParameter}.

\FloatBarrier

\subsection{Data Storage}


The data is stored in the \href{https://my.pcloud.com/}{pCloud} and contains process force measurement data, cutting edge radii scans, pictures of chip geometries and etched chips, see the respective links in table \ref{Tab:Datenablage}.

\begin{table}[h]
	\small
	\begin{center}
		\begin{tabular}{l | c | c}
			Data & Saving location & Comment\\
			\hline
			Process forces & \href{https://e1.pcloud.link/publink/show?code=kZqCC4ZnPc3kxsujyVIFiCN1bma05xPpv9V}{pCloud} & data format: time [s], $F_p [N/mm]$, $F_f [N/mm]$, $F_c [N/mm]$\\
			Cutting edge scans & \href{https://e1.pcloud.link/publink/show?code=kZDCC4Z0pbIhO0xClV0wqno2S6PFBz5zU9y}{pCloud} & new inserts, informations in CutPositionsOnCuttingEdge.xls\\
			Chip images¸& \href{https://e1.pcloud.link/publink/show?code=kZtkj4ZNRuWpXaknWpstGwCLJ0M1BnLqusX}{pCloud} & overview\\
			Chip images (polished) & \href{https://e1.pcloud.link/publink/show?code=kZACC4ZRpOt6BVmye4Ery3112CWe8zpVh37}{pCloud} & microscope images\\
			Chip images (etched) & \href{https://e1.pcloud.link/publink/show?code=kZaCC4ZhRMcWypDrM5S1Lp1HIxdAuFN8pyy}{pCloud} & microscope images with chip microstructures\\
		\end{tabular}
		\caption{Data storage of the quasi-orthogonal cutting experiments.}
		\label{Tab:Datenablage}
	\end{center}
\end{table}


\FloatBarrier

\section{Conclusions}
\label{Kap:Conclusions}

Large scale dry quasi-orthogonal cutting experiments have been conducted for \Titan and Ck45. Prior to the cutting tests a material characterization was conducted revealing:

\begin{itemize}
	\item uniform hardness profiles in hardness measurements, 
	\item uniform grain structures in etched samples,
	\item average grain diameters in the order of $D_g \approx 9..10 \mu m$ for \Titan and $D_g \approx 14..18 \mu m$ for Ck45 and
	\item yield strengths in the order of $\sigma_y = 867MPa$ for \Titan and $\sigma_y = 392MPa$ for Ck45.
\end{itemize}


Johnson-Cook flow stress model coefficients were given for the work hardening and strain rate sensitivity part, where the latter is valid only for very low strain rates. Since the tensile tests were conducted at room temperature, the temperature dependent part of the JC-law was not fitted.

Each cutting experiment was performed with a pristine cutting edge such that wear effects on the measured process forces could mostly be neglected. Since the cutting edge radius influences the process forces, every cutting edge was measured prior to the cutting tests, such that the average cutting edge radius could be retrieved at the respective cutting position.

For \TitanPunkt, the influence of the cutting edge radius on the measured process forces could be shown. Ploughing effects on the process forces were visible towards lower feeds. A friction coefficient of about $\mu_{fr} \approx 0.3$ was estimated based on the process forces depending on the feed rates. Chip segmentations were found at all tested combinations of feeds and cutting speeds.

In the Ck45 cutting experiments a direct correlation of the cutting edge radius with the process forces is not visible. This is induced due to changing physics, e.g. dynamic strain aging, of the chip formation at different feeds and cutting speeds, where at lower values evidence for built-up edge formation was found on the chips and towards higher values chip segmentations occured. The friction is not just depending on the feed, but also on the cutting speed where at medium cutting speeds the friction is higher than at very low and very high cutting speeds.

The measurement of chip thicknesses came along with scatter of the measured values, potential reasons for this scatter are:

\begin{itemize}
	\item the manual measurement procedure,
	\item misalignments of chips in the embedding procedure because the chips are curled not only in feed direction but also due to the cylinder radius and
	\item low stiffness of thinner chips which could lead to skewing and therefore larger effective cross-sections after embedding.
\end{itemize}

The low stiffness of thinner chips prevented as well from the large scale evaluation of chip curling radii. Chip curling with the cylinder radius can be prevented by using a planing setup which is the best approximation to orthogonal cutting at the expense of lower maximum cutting speeds in the order of up to $v_c=30m/min$.

%
%

\section{Acknowledgements}
\label{Kap:Acknowledgements}

The authors would hereby like to thank Adrian Fuhrer, Igor Podjanin, Knut Krieger, Matthias R\"othlin, Mohamadreza Afrasiabi, Fabian Kneub\"uhler, Moritz Wiessner, Mikhail Klyuev, Sandro Wigger, Albert Weber, Christoph Baumgart, Lukas Seeholzer, the IVP institute at ETH Z\"urich, Dr. Thomas Tancogne-Dejean (ETH) for the help with the EBSD images and the Swiss National Science Foundation (SNF) for the financial support under Grant No. 200021-149436.

\appendix

\section{Test Overview}
\label{Kap:Anhang_Versuchsuebersicht}

Tables \ref{Tab:Versuchsuebersicht_Ti6Al4V} and \ref{Tab:Versuchsuebersicht_Ck45} contain the cutting experiment numbers sorted according to feed (columns) and cutting speeds (rows) for \Titan and Ck45, respectively. The cutting experiment numbers can be used to find more information of the respective experiment in tables \ref{Tab:TestResults_Ti6Al4Vkpl} and \ref{Tab:TestResults_Ck45kpl}. The colouring of the experiment number indicates:

\begin{itemize}
	\item \textcolor{black}{black}: no chip thickness measurement available,
	\item \textcolor{orange}{orange}: chip thickness measurement available,
	\item \textcolor{green}{green}: chip thickness measurement and etched chip picture available.
\end{itemize}

\begin{table}
	\tiny
	\begin{center}
		\begin{tabular}{ p{1.1cm} | p{0.6cm}  | p{0.6cm}  | p{0.6cm} | p{0.6cm} | p{0.6cm} | p{0.6cm} | p{0.6cm} | p{0.6cm} | p{0.6cm} | p{0.6cm} | p{0.6cm} | p{0.6cm} | p{0.6cm} }
			\input{Versuchsuebersicht_vc_f_Ti6Al4V.inc}
		\end{tabular}
	\end{center}
	\caption{\TitanPunkt: Overview of process parameter combinations and according test numbers.}
	\label{Tab:Versuchsuebersicht_Ti6Al4V}
\end{table}

\begin{table}
	\tiny
	\begin{center}
		\begin{tabular}{ p{1.1cm} | p{0.6cm}  | p{0.6cm}  | p{0.6cm} | p{0.6cm} | p{0.6cm} | p{0.6cm} | p{0.6cm} | p{0.6cm} | p{0.6cm} | p{0.6cm} | p{0.6cm} | p{0.6cm} | p{0.6cm} }
			\input{Versuchsuebersicht_vc_f_Ck45.inc}
		\end{tabular}
	\end{center}
	\caption{Ck45: Overview of process parameter combinations and according test numbers.}
	\label{Tab:Versuchsuebersicht_Ck45}
\end{table}

\FloatBarrier

\section{Experimental Results}
\label{Kap:Anhang_Meszdaten}

The complete cutting test results are provided for \Titan (V0001-V0068, V0301-V0520) in table \ref{Tab:TestResults_Ti6Al4Vkpl} and for Ck45 (V0069-V0300) in table \ref{Tab:TestResults_Ck45kpl} below. Blue coloured experiment numbers contain a link to a high-speed camera video of that test. The process forces $F_c$ and $F_f$ are normalized to a cutting width of $w=1mm$ and are given together with their respective standard deviations $\sigma_{F_c}$ and $\sigma_{F_f}$. The next two columns contain the averaged cutting edge radius $r_n$ and its standard deviation $\sigma_{r_n}$, followed by the average chip thickness $h_{avg}$ and its standard deviation $\sigma_{h_{avg}}$ (if more than one chip is evaluated), the cutting distance $l_{cut}$ and the temperature of the chip $T_{chip}$ derived from the tempering color (only Ck45). The last two columns contain a qualitative statement of the tool wear and \textsc{Status} judges the quality of the process forces measurement. Tests labelled with \textcolor{green}{ok} are without any objections. Tests labelled \textcolor{orange}{short} have to be treated with care as such experiments should be repeated with a longer cutting distance in order for the process forces to reach a stable value. If however the standard deviations in the process forces of such tests are small, they still can be considered as valid. Results from tests labelled \textcolor{orange}{questionable} or \textcolor{orange}{initially stable} should not be used for parameter identifications while tests labelled with \textcolor{red}{saturation} ran into the amplifier limits and are therefore invalid for further use. Tests marked \textcolor{red}{instable} are most likely of insufficient quality and must not be used for parameter identifications as well. Nonetheless, invalid experiments are listed in the results tables for two reasons. First, critical process parameters are known then. Second, the generated chips can be still used for evaluation at these process conditions.


{\tiny

}

\FloatBarrier

\section{Kienzle Parameters}
\label{Kap:Anhang_KienzleParameter}

The Kienzle coefficients are given as a function of cutting speed for \Titan in table \ref{Tab:Kienzle_Ti6Al4V} and for Ck45 in \ref{Tab:Kienzle_Ck45}.

\begin{table}
	\tiny
	\begin{center}
	\end{center}
	\caption{\TitanPunkt: Kienzle coefficients for feed and cutting force estimations.}
	\label{Tab:Kienzle_Ti6Al4V}
\end{table}

\begin{table}
	\tiny
	\begin{center}
	\end{center}
	\caption{Ck45: Kienzle coefficients for feed and cutting force estimations.}
	\label{Tab:Kienzle_Ck45}
\end{table}

\FloatBarrier








\section*{References}


\bibliographystyle{plainnat}
\addcontentsline{toc}{section}{\refname}\bibliography{Literatur.bib}

\end{document}

%% file: Versuchsuebersicht_vc_f_Ti6Al4V.inc
$f [mm/rev]$ / $v_c [m/min]$ & 0.01 & 0.02 & 0.03 & 0.04 & 0.06 & 0.08 & 0.1 & 0.12 & 0.15 & 0.2 & 0.4\\
\hline
10.0 &  & V0301, \textcolor{green}{V0302}, V0303 &  &  & \textcolor{green}{V0304}, V0305, V0306 &  & V0307, V0308, \textcolor{green}{V0309}, V0319 &  &  &  & \\
\hline
11.0 & \textcolor{green}{V0001}, \textcolor{green}{V0003}, \textcolor{orange}{V0014} &  &  &  &  &  & \textcolor{green}{V0005}, \textcolor{green}{V0007}, \textcolor{orange}{V0016}, \textcolor{orange}{V0018} &  &  & \textcolor{green}{V0010}, \textcolor{green}{V0012}, \textcolor{orange}{V0020} & \\
\hline
13.0 & \textcolor{green}{V0002}, \textcolor{orange}{V0013}, \textcolor{orange}{V0015} &  &  &  &  &  & \textcolor{green}{V0004}, \textcolor{green}{V0006}, \textcolor{green}{V0008}, \textcolor{green}{V0009}, \textcolor{orange}{V0017} &  &  & \textcolor{green}{V0011}, \textcolor{orange}{V0019}, \textcolor{orange}{V0021} & \\
\hline
20.0 & \textcolor{green}{V0320}, V0321, V0322 & V0356, V0357, \textcolor{green}{V0358} &  & V0323, V0324, \textcolor{green}{V0325} & \textcolor{green}{V0374}, V0375, V0376, V0392 & V0359, \textcolor{green}{V0360}, V0361 & V0326, V0327, \textcolor{green}{V0328} & \textcolor{green}{V0377}, V0378, V0379 & V0362, \textcolor{green}{V0363}, V0364 & \textcolor{green}{V0380}, V0381, \textcolor{green}{V0382} & \\
\hline
40.0 & V0338, V0339, \textcolor{green}{V0340} &  &  & V0341, V0342, \textcolor{green}{V0343} & V0480, \textcolor{green}{V0481}, V0482 &  & V0344, V0345, \textcolor{green}{V0346} &  & V0483, \textcolor{green}{V0484}, V0485 & V0486, V0487, \textcolor{green}{V0488} & \\
\hline
50.0 &  & V0310, \textcolor{green}{V0311}, V0312 &  & V0393, \textcolor{green}{V0394}, V0395 & V0313, V0314, \textcolor{green}{V0315} & V0396, \textcolor{green}{V0397}, V0398 & V0316, V0317, \textcolor{green}{V0318} &  & V0399, \textcolor{green}{V0400}, V0401 & \textcolor{green}{V0414}, V0415, V0416 & \\
\hline
52.0 &  & V0417, V0418, \textcolor{green}{V0419} &  &  &  & \textcolor{green}{V0420}, V0421, V0422 &  & \textcolor{green}{V0423}, V0424, V0425 &  &  & \\
\hline
60.0 &  & V0435, V0436, \textcolor{green}{V0437}, V0444, \textcolor{green}{V0445} &  &  &  &  & V0438, V0439, \textcolor{green}{V0440} &  &  & V0441, \textcolor{green}{V0442}, V0443 & \\
\hline
70.0 &  &  &  & \textcolor{green}{V0402}, V0403, V0404 & \textcolor{green}{V0383}, V0384, V0385 & \textcolor{green}{V0405}, V0406, V0407 & V0411, \textcolor{green}{V0412}, V0413 & \textcolor{green}{V0386}, V0387, V0388 & V0408, V0409, \textcolor{green}{V0410} & V0389, \textcolor{green}{V0390}, V0391 & \\
\hline
74.0 & \textcolor{green}{V0022}, \textcolor{green}{V0024} &  &  &  &  &  & \textcolor{orange}{V0026} &  &  & \textcolor{orange}{V0028}, \textcolor{orange}{V0030} & \\
\hline
80.0 &  & V0446, V0453, \textcolor{green}{V0454} &  &  &  &  & V0447, \textcolor{green}{V0448}, V0449 &  &  & V0450, V0451, \textcolor{green}{V0452} & \\
\hline
89.0 & \textcolor{green}{V0023} &  &  &  &  &  & \textcolor{orange}{V0025}, \textcolor{orange}{V0027} &  &  & \textcolor{orange}{V0029} & \\
\hline
100.0 & V0329, V0330, \textcolor{green}{V0331} & \textcolor{green}{V0365}, V0366, V0367 &  & V0332, \textcolor{green}{V0333}, V0334 & V0498, \textcolor{green}{V0499}, V0500 & \textcolor{green}{V0368}, V0369, V0370 & V0335, \textcolor{green}{V0336}, V0337 & \textcolor{green}{V0501}, V0502, V0503 & V0371, V0372, \textcolor{green}{V0373} & V0504, \textcolor{green}{V0505}, V0506 & \\
\hline
125.0 & V0347, \textcolor{green}{V0348}, V0349 &  &  & \textcolor{green}{V0350}, V0351, V0352 & V0489, \textcolor{green}{V0490}, \textcolor{green}{V0491} &  & \textcolor{green}{V0353}, V0354, V0355 &  & V0492, V0493, \textcolor{green}{V0494} & V0495, \textcolor{green}{V0496}, V0497 & \\
\hline
150.0 &  & \textcolor{green}{V0516}, V0517, V0518 & \textcolor{green}{V0519}, V0520 &  & V0507, \textcolor{green}{V0508}, V0509 &  & \textcolor{green}{V0455}, V0456, V0457 & \textcolor{green}{V0510}, V0511, V0512 &  & V0513, \textcolor{green}{V0514}, V0515 & \\
\hline
159.0 & \textcolor{green}{V0032} &  &  &  &  &  & \textcolor{orange}{V0034} &  &  & \textcolor{orange}{V0037}, \textcolor{orange}{V0039} & \textcolor{orange}{V0041}\\
\hline
162.0 &  & V0426, \textcolor{green}{V0427}, V0428 &  &  &  & \textcolor{green}{V0429}, V0430, V0431 &  & V0432, \textcolor{green}{V0433}, V0434 &  &  & \\
\hline
191.0 & \textcolor{green}{V0031}, \textcolor{orange}{V0033} &  &  &  &  &  & \textcolor{orange}{V0035} &  &  & \textcolor{orange}{V0038} & \textcolor{orange}{V0040}, \textcolor{orange}{V0042}\\
\hline
200.0 &  &  &  &  &  &  & V0458, \textcolor{green}{V0459}, V0460 &  &  &  & \\
\hline
212.0 & \textcolor{green}{V0043}, \textcolor{green}{V0045} &  &  &  &  &  & \textcolor{orange}{V0047}, \textcolor{orange}{V0049} &  &  & \textcolor{orange}{V0051} & \textcolor{orange}{V0053}, \textcolor{orange}{V0055}\\
\hline
254.0 & \textcolor{green}{V0044}, \textcolor{orange}{V0046} &  &  &  &  &  & \textcolor{orange}{V0048} &  &  & \textcolor{orange}{V0050}, \textcolor{orange}{V0052}, \textcolor{orange}{V0068} & \textcolor{orange}{V0054}\\
\hline
318.0 & \textcolor{green}{V0057} &  &  &  &  &  & \textcolor{green}{V0059}, \textcolor{green}{V0061} &  &  & \textcolor{green}{V0063} & \textcolor{orange}{V0065}, \textcolor{orange}{V0067}\\
\hline
381.0 & \textcolor{green}{V0056}, \textcolor{green}{V0058} &  &  &  &  &  & \textcolor{green}{V0060} &  &  & \textcolor{green}{V0062}, \textcolor{green}{V0064} & \textcolor{orange}{V0066}\\
\hline
400.0 & \textcolor{green}{V0461}, V0462, V0463 & V0464, V0465, \textcolor{green}{V0466} & V0467, \textcolor{green}{V0468}, V0469 & V0470, \textcolor{green}{V0471}, V0472 &  &  &  &  &  &  & \\
\hline
500.0 & V0473, \textcolor{green}{V0474}, V0475 & V0476, V0477, \textcolor{green}{V0478} &  &  &  &  & \textcolor{green}{V0479} &  &  &  & \\

%% file: Versuchsuebersicht_vc_f_Ck45.inc
$f [mm/rev]$ / $v_c [m/min]$ & 0.01 & 0.02 & 0.04 & 0.06 & 0.08 & 0.1 & 0.12 & 0.15 & 0.18 & 0.2 & 0.22 & 0.4 & 1.0\\
\hline
10.0 & V0069, \textcolor{green}{V0070}, V0071, V0212, \textcolor{green}{V0213}, V0214 & \textcolor{green}{V0176}, \textcolor{green}{V0177}, V0178 & \textcolor{green}{V0215}, V0216, V0217 & \textcolor{green}{V0179}, V0180, V0181 & V0218, \textcolor{green}{V0219}, \textcolor{green}{V0220} & \textcolor{green}{V0072}, \textcolor{green}{V0073}, \textcolor{green}{V0074} &  & V0182, V0183, \textcolor{green}{V0184} &  & \textcolor{green}{V0075}, \textcolor{green}{V0076}, \textcolor{green}{V0077}, \textcolor{green}{V0078} &  &  & \\
\hline
30.0 &  & V0102, V0103, \textcolor{green}{V0104} & V0120, \textcolor{green}{V0121}, V0122 & V0105, \textcolor{green}{V0106}, V0107 &  & V0123, V0124, \textcolor{green}{V0125} &  & \textcolor{orange}{V0108}, \textcolor{orange}{V0109}, \textcolor{orange}{V0110} &  & V0126, \textcolor{green}{V0127}, V0128 &  &  & V0138\\
\hline
50.0 & V0252, \textcolor{green}{V0253}, V0254 & \textcolor{green}{V0139}, \textcolor{green}{V0140}, V0141 & V0157, \textcolor{green}{V0158}, V0159 & V0142, \textcolor{green}{V0143}, V0144 & \textcolor{green}{V0255}, \textcolor{green}{V0256}, V0257 & V0160, V0161, \textcolor{green}{V0162} & V0258, \textcolor{green}{V0259}, V0260 & V0145, \textcolor{green}{V0146}, V0147 & V0270, \textcolor{green}{V0271}, V0272, V0273 & V0163, \textcolor{green}{V0164}, \textcolor{green}{V0165}, V0295, \textcolor{green}{V0296}, V0297 & V0274, \textcolor{green}{V0275}, V0276 &  & \\
\hline
70.0 & \textcolor{green}{V0079}, V0080, \textcolor{green}{V0081} & V0185, \textcolor{green}{V0186}, V0187 &  & V0188, \textcolor{green}{V0189}, V0190 &  & V0082, V0083, \textcolor{green}{V0084}, \textcolor{green}{V0085} &  & \textcolor{green}{V0191}, V0192, V0193 &  & V0086, \textcolor{green}{V0087}, \textcolor{green}{V0088}, V0089 &  &  & \\
\hline
100.0 &  & V0111, \textcolor{green}{V0112}, \textcolor{green}{V0113} & \textcolor{green}{V0129}, V0130, V0131 & V0114, V0115, \textcolor{green}{V0116} &  & V0132, \textcolor{green}{V0133}, V0134 &  & \textcolor{green}{V0117}, V0118, V0119 &  & V0135, V0136, \textcolor{green}{V0137} &  &  & \\
\hline
150.0 & \textcolor{green}{V0090}, \textcolor{green}{V0091}, \textcolor{green}{V0092} & V0148, V0149, \textcolor{green}{V0150}, \textcolor{green}{V0175} & V0166, \textcolor{green}{V0167}, \textcolor{green}{V0168} & \textcolor{green}{V0151}, V0152, V0153 &  & \textcolor{green}{V0093}, \textcolor{green}{V0094}, \textcolor{green}{V0095}, \textcolor{green}{V0169}, V0170, \textcolor{green}{V0171} &  & \textcolor{green}{V0154}, V0155, V0156 &  & \textcolor{green}{V0096}, \textcolor{green}{V0097}, \textcolor{green}{V0098}, V0172, \textcolor{green}{V0173}, \textcolor{green}{V0174} &  & \textcolor{green}{V0099}, \textcolor{green}{V0100}, \textcolor{green}{V0101} & \\
\hline
200.0 & \textcolor{green}{V0194}, V0195, V0196 & V0277, \textcolor{green}{V0278}, V0279 & V0197, \textcolor{green}{V0198}, V0199 & \textcolor{green}{V0280}, V0281, V0282 & \textcolor{green}{V0200}, \textcolor{green}{V0201}, V0202 & V0283, V0284, \textcolor{green}{V0285} &  &  &  &  &  &  & \\
\hline
250.0 & \textcolor{green}{V0261}, V0262, V0263 & V0230, \textcolor{green}{V0231}, V0232 &  & V0233, \textcolor{green}{V0234}, V0235 & V0264, V0265, \textcolor{green}{V0266} & \textcolor{green}{V0236}, V0237, V0238 & V0267, V0268, \textcolor{green}{V0269} &  &  &  &  &  & \\
\hline
300.0 & V0203, \textcolor{green}{V0204}, \textcolor{green}{V0205} &  & V0206, \textcolor{green}{V0207}, V0208 &  & V0209, \textcolor{green}{V0210}, V0211 &  &  &  &  &  &  &  & \\
\hline
350.0 & V0221, V0222, \textcolor{green}{V0223} & \textcolor{green}{V0239}, V0240, V0241 & V0224, \textcolor{green}{V0225}, V0226 & V0242, V0243, \textcolor{green}{V0244} & V0227, \textcolor{green}{V0228}, V0229 & \textcolor{green}{V0245}, V0246, V0247 &  & V0248, \textcolor{green}{V0249}, V0250 &  &  &  &  & \\
\hline
450.0 &  & \textcolor{green}{V0286}, V0287, V0288 &  & \textcolor{green}{V0289}, V0290, V0291 &  & V0292, \textcolor{green}{V0293}, V0294 &  &  &  &  &  &  & \\
\hline
500.0 &  & \textcolor{green}{V0251} &  &  &  & V0298, \textcolor{green}{V0299}, \textcolor{green}{V0300} &  &  &  &  &  &  & \\